\documentclass[preprint]{aastex}
\input epsf
\newcommand{\reff}{\mbox{$R_{\rm e}$}}

\newcommand{\ishape}{{\tt ishape}}
\newcommand{\tinytim}{{\tt TinyTim}}
\newcommand{\dsmall}{\mbox{$\Delta\rm m_{2\rightarrow5}$}}
\newcommand{\dlarge}{\mbox{$\Delta\rm m_{3\rightarrow5}$}}
\newcommand{\mto}{\mbox{$M_{\rm TO}(V)$}}
\newcommand{\vi}{\mbox{$V\!-\!I$}}
\newcommand{\viz}{\mbox{$(V\!-\!I)_0$}}
\begin{document}
\title{HST observations of star clusters in NGC 1023: Evidence for
 three cluster populations?
 \footnote{Based on observations with the NASA/ESA Hubble Space 
 Telescope, obtained at the Space Telescope Science Institute, which is 
 operated by the Association of Universities for Research in Astronomy, 
 Inc. under NASA contract No.  NAS5-26555.}
}
\author{S{\o}ren S. Larsen and Jean P. Brodie}
\affil{UC Observatories / Lick Observatory, University of California,
       Santa Cruz, CA 95064, USA}
\email{soeren@ucolick.org and brodie@ucolick.org}

\begin{abstract}
  Using HST images we have carried out a study of cluster populations
in the nearby S0 galaxy NGC~1023.  In two WFPC2 pointings we have 
identified 221 cluster candidates.  The small distance ($\sim9$ Mpc) combined
with deep F555W and F814W images allows us to reach about two magnitudes 
below the expected turn-over of the globular cluster luminosity function. 
NGC~1023 appears to contain at least three identifiable cluster 
populations: the brighter clusters show a clearly bimodal color distribution 
with peaks at $\viz = 0.92$ and at $\viz = 1.15$ and in addition there are
a number of fainter, more extended objects with predominantly red colors. 
Among the brighter clusters, we find that the blue clusters
have somewhat larger sizes than the red ones with mean effective radii
of $\reff \sim 2$ and $\reff \sim 1.7$ pc, respectively. These clusters
have luminosity functions (LFs) and sizes consistent with what is observed 
for globular clusters in other galaxies. Fitting Gaussians to the LFs of the
blue and red compact clusters we find turn-over magnitudes of 
$M_{\rm TO}({\rm blue})=-7.58^{-7.72}_{-7.36}$ and 
$M_{\rm TO}({\rm red})=-7.37^{-7.50}_{-7.09}$ in $V$ and dispersions of
$\sigma_V({\rm blue})=1.12_{1.03}^{1.33}$ and 
$\sigma_V({\rm red})=0.97_{0.89}^{1.25}$. 
The fainter, more extended clusters have effective radii up to 
$\reff \sim 10-15$ pc and their LF appears to rise at least down to 
$M_V \sim -6$, few of them being brighter than $M_V = -7$. We suggest that 
these fainter objects may have a formation history distinct from 
that of the brighter GCs.
\end{abstract}

\keywords{galaxies: elliptical and lenticular, cD ---
          galaxies: star clusters ---
	  galaxies: individual (NGC~1023)}

\section{Introduction}

  During the last two decades a large amount of information about globular
cluster systems (GCSs) in other galaxies has been collected. One of the 
most remarkable findings is the apparent constancy of the globular cluster 
luminosity function (GCLF), which is nearly always well fitted by a 
Gaussian-like function with a turn-over magnitude
\mbox{($M_{\rm TO}$)} at $M_V\sim-7.5$ 
plus/minus a few tenths of a magnitude at most \citep{har99} and a 
dispersion of $\sigma_V \sim 1.2$. Thus, the GCLF has gained much potential 
as a distance indicator. Note that the GCLF is Gaussian in {\it magnitude} 
units. When plotted in luminosity units it is well represented as a 
composite of two power-laws with a change in slope at a luminosity
roughly corresponding to the turn-over magnitude \citep{har94}.

  Another intriguing discovery is that many GCSs have bimodal color 
distributions, generally thought to represent GC populations of
different metallicities and/or ages. Recent studies \citep{geb99,kun99a} 
have shown this to be a very common phenomenon, providing evidence that 
many galaxies must have had several distinct episodes of star formation.  
The GCSs of a few galaxies such as M87 and NGC~1399 have even been suggested 
to contain three metallicity peaks \citep{lee93}.  A number of scenarios
have been suggested to account for the presence of several cluster
populations in galaxies. These include formation of the metal-rich population
by gas-rich mergers \citep{ash92}, in late stages of a multi-phase
collapse \citep{for97}, or accretion of the {\it metal-poor} population
in giant ellipticals from less luminous galaxies \citep{cot98,hil99}. However,
no single theory can be said to satisfactorily account for all observed
properties of globular cluster systems in different galaxies and the
final answer may possibly lie in a combination of the above and/or other
scenarios. A thorough discussion of various formation scenarios is given
in \citet{van00}.

  In spite of the advances and the vast amount of data collected, many
fundamental questions about the properties of GCSs still remain to be
answered. One of the most elusive problems has been to explain the
constancy of the GCLF turn-over and it is still debated whether the 
LF of old GCSs can be explained by dynamical 
evolution from the power-law luminosity distribution that is observed in 
{\it young} cluster systems in mergers and starburst galaxies
\citep{elm97,whi99,zha99,fri99}.
In this respect it is important to study the faint wing of the GCLF 
($M_V > -7.5$) where the effects of dynamical evolution would be most 
pronounced. For most galaxies this remains a difficult task even 
with HST photometry, so although the turn-over magnitude may indeed be 
universal, much less is known about whether or not the shape of the GCLF is 
really constant, especially to fainter magnitudes.

  In this paper we report the results of a study of star clusters in
the nearby SB0 galaxy NGC~1023. The generic term ``star clusters'' is here
preferred to ``globular clusters'' for reasons that will become apparent
later.  We adopt a distance modulus of $m-M = 29.97\pm0.14$, measured 
using planetary nebulae \citep{cia91} and consistent with the HST Cepheid 
distance of another member of the same group, NGC~925 
($m-M = 29.84\pm0.16$, Silbermann et al.~1996).
The GCLF turn-over is thus expected to be at 
$V \sim 22.5$, bright enough that we can reach at least two magnitudes below 
it with our HST data and probe the faint end of the GCLF. At the distance
of NGC~1023,  1\arcsec\ corresponds to 48 pc, so objects with sizes of a 
few pc (typical for globular clusters in the Milky Way and other
galaxies) will appear relatively well resolved and good estimates of their 
sizes can be obtained.
  For the interstellar absorption towards NGC~1023 we adopt 
$E(\bv ) = 0.061$ \citep{sch98}, corresponding to 
$E(\vi ) = 0.10$ and $A_V = 0.19$ \citep{car89}. This is very 
similar to the $E(\bv ) = 0.06$ found by \citet{kin82}, as well as 
to the $A_B$ value of 0.27 (equivalent to $E(\bv ) = 0.066$) given in the 
RC3 catalogue.

  NGC~1023 is included in Arp's {\it Atlas of Peculiar Galaxies}
\citep{arp66} because of a small eastern companion, NGC~1023A.
On optical images NGC~1023A looks like an extension of the main galaxy
to the E, but its \ub\ color is noticeably bluer than that of NGC~1023 itself 
and implies that active star formation has 
proceeded in the companion until $\sim 200$ Myr ago \citep{cap86}.
NGC~1023 is known to be rich in HI gas, possibly associated with the 
companion, and HI maps \citep{san84} show that the highest concentration of 
HI gas in the NGC~1023 system is indeed coincident with the location of 
NGC~1023A. The radial velocity of the HI gas near NGC~1023A is also 
consistent with the gas being associated with NGC~1023A \citep{cap86}.
Much of the remaining gas forms a ring-like structure around NGC~1023. 
Several other dwarf galaxies are known near NGC~1023 \citep{dav84} and 
it seems plausible that the ongoing accretion of NGC~1023A may just be 
the most recent of several similar events that have taken place in the 
past. Such interactions could have had significant impacts on the
star- and cluster formation history of NGC~1023 itself.

\section{Data}

  WFPC2 images in the F555W and F814W filters were obtained in Cycle 6 for 
two pointings in NGC~1023, one with the Planetary Camera (PC) centered on the
galaxy nucleus and another further to the East (Fig.~\ref{fig:pointings}). 
For both pointings the total integration time in each filter was 2400 sec, 
split into two integrations.

  The two exposures for each pointing in each of the F555W and F814W 
filters were combined using the IRAF {\tt imcombine} task with the 
{\tt reject} option set to {\bf crreject}, eliminating Cosmic Ray events.
Prior to photometric analysis, we removed large-scale background variations 
by a two-step process:
First, the images were smoothed with a $15\times 15$ box median filter
and the smoothed images were then subtracted from the original set.
Objects were detected on this first preliminary set of 
background-subtracted images using the {\tt daofind} task in the
DAOPHOT package \citep{ste87} and the detected objects were subtracted
from the original images using the \ishape\ algorithm \citep{lar99}.
\ishape\ is capable of dealing with extended objects of varying sizes,
in contrast to e.g.\ the {\tt substar} task in DAOPHOT. 
The object-subtracted images were smoothed as before, resulting in a
set of images with all small-scale variations removed, which were then
subtracted from the original images to produce our final set of
background-subtracted images for further analysis.

  Input object lists for photometry were produced by running {\tt daofind} 
on the background-subtracted images in both of the F555W and F814W bands
and matching the two lists. As an additional selection criterion, the
background noise was measured directly on the image in an annulus around
each object and only objects with a S/N $>3$ in both F555W and F814W within
an aperture radius of 2 pixels were accepted. In this way we accounted for
the varying background noise within the image.

Photometry was then obtained using the 
{\tt phot} task in DAOPHOT. Because the {\it difference} between aperture
corrections in different filters is quite independent of object
size \citep[see also Sect.~\ref{sec:apc}]{hol96,lar99},
we used a small aperture radius ($r=2$ pixels)
for \vi\ colors.  Selection of the aperture for $V$ magnitudes represented 
a more difficult compromise between eliminating systematic and random 
errors, but we found an $r=3$ pixels aperture to be a reasonable choice. The 
calibration to $V$ and $I$ band magnitudes was done following the procedure 
described in \citet{hol95}.

  A second run of \ishape\ was also performed in order to get intrinsic 
sizes for all objects detected in the second pass. The HST PSF was
modeled using the \tinytim\ simulator \citep{kri97} and King 
profiles with concentration parameter $c=30$ were assumed for the cluster 
profile models. Here $c$ is the ratio between tidal and core radius.  
\ishape\ produced estimates of the FWHM of each cluster, which were then
converted to half-light radii. The \ishape\ modeling also included 
convolution with the WFPC2 diffusion kernel. 

  In order to test the reliability of the cluster sizes we added artificial
objects, generated by convolving the \tinytim\ PSF with King profiles
of various FWHM values, to the NGC~1023 images. The sizes of the artificial
objects were then remeasured using \ishape . Repeating this experiment for
artificial objects of different magnitudes, we found that the input cluster 
sizes could be reproduced with an accuracy of about $\pm$ 30\% down to $V=24$. 

  As a further check of \ishape\ we compared our FWHM values with the 
$\Delta_{0.5-3}$ index, originally defined by \citet{sch96} and
since then used by many authors to measure sizes of extended objects on
HST images. Fig.~\ref{fig:delta} shows the $\Delta_{0.5-3}$ values versus
FWHM estimates by \ishape\ for objects with $V<24$. A quite good, although
not linear correlation exists, giving confidence to the cluster sizes
measured by \ishape . We also checked the \ishape\ cluster sizes from
F555W and F814W images against each other, again providing excellent
agreement.

  Because of possible systematic differences between aperture corrections
and object sizes measured on the PC and WF camera chips, and because of 
the very high surface
brightness in the central parts of NGC~1023, we only considered objects 
in the WF chips.  Inspection by eye revealed at most $\sim 10$ cluster 
candidates on the PC chip of the central pointing down to the 
detection limit.

\subsection{Aperture corrections}
\label{sec:apc}

  In order to measure \dsmall\ and \dlarge\ aperture corrections for
extended sources from the $r=2$ and $r=3$ pixels apertures to the 
\citet{hol95} $r=5$ pixels reference aperture, we convolved the \tinytim\ 
PSF with King $c=30$ and MOFFAT15 \citep{lar99} profiles of different 
FWHM values. The aperture corrections were then obtained by carrying 
out aperture photometry on the convolved model profiles, measuring the 
differences in the fluxes through various apertures. The experiment was 
carried out for a number of different FWHM values for the model profiles, 
ranging between 0.1 and 2.0 pixels (0.5 -- 10 pc at the assumed distance 
of NGC~1023). The FWHM can be converted to effective radii (\reff) by 
multiplying with 1.48 for the King $c=30$ profiles and by 1.13 
for the MOFFAT15 profiles.

  The aperture corrections measured in this way are listed in 
Table~\ref{tab:apc} for the F555W band and for the difference F555W--F814W.
In addition to the \dsmall\ and \dlarge\ corrections
between the $r=5$ reference aperture and the apertures used for
our photometry, Table~\ref{tab:apc} also lists aperture corrections
from $r=5$ to $r=30$ pixels. 

  Not surprisingly, the aperture corrections in F555W depend strongly
on object size. For point sources the \dlarge(F555W) correction 
is $-0.060$ mag, increasing to $-0.15$ for a King $c=30$
profile with FWHM=0.50 pixels and reaching $-0.43$ mag for FWHM=2.0 pixels.
For a given FWHM, the aperture corrections for the MOFFAT15 profiles
are somewhat smaller than for the King $c=30$ profiles because a larger 
fraction of the flux is contained in the wings of the King $c=30$ profile. 
For a King profile with a smaller concentration parameter, the aperture
corrections would be smaller for any given FWHM. 

  A correction of $-0.1$ mag from the $r=5$ pixels aperture to infinity
is implicit in the standard calibration procedure, in good agreement with 
our results for point sources listed in Table~\ref{tab:apc}.  According to 
\citet{hol95}, this correction is nearly independent of the choice of filter, 
small changes in telescope focus etc. However, for extended objects a larger
fraction of the light falls outside the $r=5$ aperture, amounting to an
error of about $-0.4$ mag for a FWHM of 2 pixels ($\reff = 14$ pc) if
no correction is applied.

  Although the aperture corrections in one filter depend strongly on
object size, {\it colors} can be measured with much better precision.
For point sources we find an aperture correction \dsmall(F555W--F814W) 
of 0.031 mag for the \vi\ colors, in good agreement with other authors 
\citep{puz99,whi97}.  The correction depends only weakly on the object size, 
implying that systematic errors in \vi\ color are small ($\la 0.01$ mag) 
even for objects with effective radii up to $\sim 15$ pc. This remains true
also when the correction from $r=5$ to infinity is considered, which is
constant within the uncertainties. The \vi\ aperture corrections from 
$r=5$ to $r=30$ given in Table~\ref{tab:apc} are, strictly speaking, 
only valid for the idealized case where the background level can be 
determined with high precision. In practice, small uncertainties in the 
background determination and in the object profiles will generally 
overshadow the small changes in the \vi\ aperture corrections from 
$r=5$ to infinity as a function of object size.

  Considering that Milky Way globular clusters typically have $\reff 
\sim 3$ pc \citep{har96}, we adopted an \dlarge(F555W) aperture correction of 
$-0.14$ mag. This aperture correction is probably accurate to about
0.1 mag for objects with intrinsic $\reff \la 5$ pc, but larger objects will
be systematically too faint by up to several tenths of a magnitude. The total
$V$ error for an $\reff=14$ pc object will amount to about 0.7 mag.  For 
\vi\ we use a correction of 0.026, which should lead to systematic errors 
of no more than about 0.01 mag or so.  

\subsection{Completeness}
\label{sec:compl}

  Because of the relatively small distance, clusters in NGC~1023 appear
quite well resolved on our HST images. This requires that special
attention be paid to completeness corrections and their dependence on
the angular extent of the objects. We carried out completeness tests
by adding synthetically generated objects to the science frames and
checking how many of the artificial objects were recovered by a
subsequent photometry run as a function of magnitude. The tests were 
repeated for a number of different object sizes, using the \tinytim\ PSF 
convolved with King profiles of FWHM ranging between 0.25 and 
2 pixels, corresponding to $\reff = 1.8 - 14.2$ pc.

  The results of the completeness tests are shown in Fig.~\ref{fig:cplot}.
The 50\% completeness limit for {\it point sources} (i.e., the pure
\tinytim\ PSF) is at $V \approx 26.2$, but for more extended objects
the curves shift rapidly to the left in the diagram, indicating that the
50\% limit for objects with $\reff = 14$ pc is at $V \approx 24.5$,
about two magnitudes brighter than for point sources. 

  The curves in Fig.~\ref{fig:cplot} represent the {\it average} completeness
functions for each field. However, the completeness also depends on the 
background level. Because the background level increases strongly towards
the center of NGC~1023, a radial dependence may be expected for the
completeness functions. In order to test this, we performed another set
of completeness tests by adding artificial objects to a number of purely 
synthetic images with different background levels, including 
photon shot noise as well as a contribution from ``read-out noise'' in 
the synthetic images. This 
allowed us to check how the completeness functions depend on the background 
level and, by comparison with the surface brightness profile of NGC~1023,
on the distance from the center of the galaxy.
Fig.~\ref{fig:cmpl2} shows the 50\% completeness limit obtained 
from these tests as a function of distance $D$ from the center of NGC~1023 
along the major axis, for objects with $\reff=2$ pc, 7 pc and 14 pc. From 
these tests we expect the 50\% completeness limit to be below $V=24$ for 
objects with $\reff < 14$ pc at $D \ga 20\arcsec$.

  Note that the $x$-axes in Figs.~\ref{fig:cplot} and \ref{fig:cmpl2} refer 
to the 
synthetic {\it input} magnitudes which, for extended objects, will be 
somewhat brighter than the magnitudes actually measured in our $r=3$ pixels 
aperture. Therefore, since extended objects are actually brighter than 
they appear, the detection of extended objects with a given {\it apparent} 
magnitude will be more efficient than what appears from the curves 
in Figs.~\ref{fig:cplot} and \ref{fig:cmpl2}.

\section{Results}

  Fig.~\ref{fig:vi_v_all} shows a $\viz , V$ color-magnitude diagram for 
all objects detected in the two WFPC2 pointings down to $V=25$, corrected 
for Galactic foreground extinction.  Below $V=25$ the photometric errors 
become very large, exceeding 0.20 in \vi\ so we do not consider objects 
fainter than this limit.  For further analysis we selected globular cluster 
(GC) candidates as objects in the color interval $0.75 < \viz < 1.40$ 
(dashed lines) and $20<V<25$, resulting in a total of 221 GC candidates.


\subsection{Colors and sizes: three cluster populations?}

  Already a casual inspection of the color-magnitude diagram in 
Fig.~\ref{fig:vi_v_all} reveals a clearly bimodal color distribution.
A homoscedastic KMM test \citep{ash94} confirms bimodality at the 
$>99.9\%$ confidence level with peaks at $\viz = 0.93$ and $\viz = 1.19$
for the sample of GC candidates, corresponding to metallicities of
[Fe/H] =$-1.5$ and [Fe/H] =$-0.6$ \citep{kis98}.  The KMM test assigns 
45.7\% of the objects to the blue peak and 54.3\% to the red one.  As a 
formal dividing line between ``blue'' and ``red'' clusters we adopt 
$\viz = 1.05$.  

  Fig.~\ref{fig:vi_fwhm} shows the sizes as a function of \viz\ 
color for cluster candidates brighter than $V=24$ (145 objects).  
Histograms of the \reff\ distributions for red ($\viz < 1.05$) and 
blue clusters ($\viz > 1.05$) are in Fig.~\ref{fig:szdist}. The size 
distribution of blue GC candidates peaks at $\reff \sim 2$ pc, while
the scatter in the size distribution for the {\it red} GC 
candidates is much larger with a peak at $\reff \sim 1$ pc and a tail 
extending up to $\reff \sim 15$ pc. For the most compact sources, the 
effective radii are not very sensitive to the specific choice of model
profile \citep{lar99}.
However, for the more extended sources the derived \reff\ values depend 
on the adopted model profiles. If we use MOFFAT15 
instead of King $c=30$ models the sizes of clusters larger than 
$\reff \sim 5$ pc are reduced by roughly 25\%, while the \reff\ values for 
the compact GCs remain essentially unaffected. The numbers given throughout 
this paper are based on King $c=30$ models.

  Fig.~\ref{fig:vi_fwhm} suggests a division of the clusters into two
groups based on their sizes, splitting at $\reff = 7$ pc or so. In the 
following, we thus distinguish between ``compact'' and ``extended'' objects as 
objects with $\reff < 7$ pc and $\reff \ge 7$ pc, respectively.  This results 
in 116 and 29 objects being assigned to the compact and 
extended groups. The exact size cut is of little importance; 
if we had used 5 pc instead of 7 pc then the number of compact and 
extended objects would have been 111 and 34. Leaving out the extended objects 
and re-running the KMM test, the color distribution is still bimodal at 
the $>99.9$\% confidence level and the two peaks are now at 
$\viz = 0.92$ and $\viz = 1.15$.  Just taking the average colors of the
blue and and red clusters gives $\viz = 0.92$ and $\viz = 1.17$, in
excellent agreement with the KMM estimates.  The relative numbers of 
blue and red objects are now 57.5\% and 42.5\%, confirming that a larger 
fraction of the red objects have $\reff > 7$ pc.

  The average sizes of blue and red compact clusters are 2.0 pc and 1.7 pc, 
while the median sizes are 1.8 and 1.1 pc, respectively. Even for objects
in the ``compact'' category, there is a significant tail up to 
$\reff \sim 5$ pc for both blue and red clusters. This is no different from 
our own Galaxy \citep{van96} and similar size distributions have also been 
noted in a number of other galaxies \citep{kun99a}.

  Fig.~\ref{fig:pops} shows a color-magnitude diagram for the same 
objects as in Fig.~\ref{fig:vi_v_all}, but now with symbol sizes 
proportional to the object sizes derived by \ishape . For a few objects
(mostly faint ones with $V>24$) the \ishape\ fits failed to converge and
hence such objects do not appear in Fig.~\ref{fig:pops}. Very compact objects 
with $\reff \la 0.2$ pc (presumably foreground stars) scatter all over 
the diagram and are not very numerous. The extended red objects
are {\it clearly fainter} than the compact red objects, most of them
having $V>23$, and appear to constitute a separate population, distinct
from the normal globular clusters in NGC~1023.

  The fact that clusters larger than about 7 pc are predominantly red
and faint is also demonstrated by Fig.~\ref{fig:sz_v} which shows the 
cluster $V$ magnitudes as a function of their sizes. Red clusters are
shown with plus markers and blue clusters with triangles. The plot
contains a few large blue objects fainter than $V=24$, but the size 
estimates become increasingly uncertain below $V=24$ and some of these 
objects may well be contaminating background galaxies. Fig.~\ref{fig:sz_v} 
is qualitatively quite similar to the corresponding plots for globular 
clusters in the Milky Way (Fig.\ 1 in \citet{van96}).  Note, however, that 
the large clusters in our own galaxy are predominantly {\it metal-poor} 
while the opposite seems to be the case in NGC~1023 (assuming that a red 
color indicates high metallicity). Also, the large clusters are much 
more numerous in NGC~1023.

\subsection{Comparison field}

  To check the contamination by foreground and background objects we
obtained photometry for a comparison field located about 2 degrees
from NGC~1023, using HST archive data (proposal ID 6254, PI: E. Groth).
The comparison data consist of $2\times1300$ s integrations in each
of the F606W and F814W filters and should thus be roughly as deep
as our NGC~1023 data, except for the wider bandpass of the F606W filter
and the lower background far from NGC~1023. Both effects will tend to
make the comparison data more complete at faint magnitudes than our 
NGC~1023 data.  

  The \vi\ color-magnitude diagram for the comparison field is shown in 
Fig.~\ref{fig:cmp}. Without applying any size cut, the comparison field 
contains 16 objects in the color-magnitude range that confines our GC 
candidates, i.e.  $0.75<\vi <1.40$ and $20<V<25$.  Since our NGC~1023 
dataset consists of two pointings, we may thus expect about 32
contaminating objects out of the 
total number of 221 GC candidates. However, most of the contaminating 
objects are quite faint and for $V<24$ we find only 5 objects in the 
comparison field in the GC candidate color range. Of these, 3 fall into 
the blue category and 2 in the red. Contamination is thus expected to
be a minor problem above $V\sim24$.  Comparison of 
Figs.~\ref{fig:vi_v_all} and \ref{fig:cmp} also shows that objects outside 
the GC color range can be easily explained as foreground and/or 
background sources.

\subsection{Spatial distributions}

  The spatial distributions of different object types are illustrated
in Fig.~\ref{fig:dist}, which shows contours of an optical image of
NGC~1023 (from the Digital Sky Survey) overlaid with our WFPC2 pointings. 
The two upper panels show the distribution of red (left) and blue (right)
compact objects. The lower left panel shows the faint extended objects
with $\reff > 7$ pc, and the lower right panel contains all objects in the
three other panels added together. Since size information is used in 
producing these plots, all objects are brighter than $V=24$.

  The spatial distributions of compact and extended objects clearly differ.
The extended sources are not nearly as concentrated towards the center of 
the galaxy and appear to be associated with the {\it disk} of NGC~1023. 
The high degree of alignment with the isophotes of NGC~1023 makes it unlikely 
that they are background sources like, for example, a distant galaxy cluster.  
Even for the extended objects, our data are expected to be more than
50\% complete for $V<24$ (Sect.~\ref{sec:compl}) except at distances smaller
than about $20\arcsec$ from the nucleus of NGC~1023 along the major axis.
This corresponds roughly to the edge of the PC chip (the small boxes
in Fig.~\ref{fig:dist}) and it is therefore clear that completeness 
effects cannot explain the observed differences in the radial distributions
of various object types.

  Among the compact GCs, we note that the distribution of red clusters 
seems to be more flattened than that of blue ones. A similar result was
found for NGC~3115 by \citet{kundu98} who suggested that the red (metal-rich)
GCs are associated with a thick-disk like population, while the blue 
(metal-poor) clusters are halo objects.

\subsection{Luminosity distributions}

  The luminosity functions for the different object types are shown in
Fig.~\ref{fig:lf}. The overall appearance of the luminosity distributions 
for compact objects appears to be consistent with a maximum at $V\sim22.3$, 
roughly as expected for a ``standard'' Gaussian GCLF.  Fig.~\ref{fig:lf_cmp}
shows the luminosity functions of red and blue objects in the comparison field.
We remind the reader that the luminosity functions in Fig.~\ref{fig:lf_cmp} 
should be multiplied by a factor of 2 (only one comparison field but two
science fields) before comparison with Fig.~\ref{fig:lf}. For both the red 
and blue objects, 
significant contamination clearly sets in below $V=24$ and many of the excess 
faint red objects in NGC~1023 (relative to the standard GCLF) below
this magnitude are likely contaminants.

  We carried out maximum-likelihood fits of Gaussian and Student's $t_5$
functions \citep{sec92} to the luminosity functions of objects with $V<24$. 
Incompleteness effects are expected to be negligible (especially for 
compact objects) above this magnitude limit, which is $\sim 1.5$ mag below 
the expected turn-over and we should thus be able to obtain quite robust
estimates of the parameters ($\mto$ and $\sigma_V$) of the GCLFs.
The results are listed in Table~\ref{tab:lf} for red and blue compact clusters
separately, for the combined sample of red and blue clusters, and for
all cluster candidates including the extended red objects. A Gaussian
fit yields $\mto = -7.37^{-7.50}_{-7.09}$ and $\mto =-7.58^{-7.72}_{-7.36}$
for the red and blue clusters, respectively, while the
dispersions are $\sigma_V = 0.97_{0.89}^{1.25}$ and $1.12_{1.03}^{1.33}$. If 
red and blue clusters are fitted simultaneously we get
$\mto = -7.48^{-7.59}_{-7.32}$ and $\sigma_V = 1.07^{1.24}_{1.00}$.
Fitting a $t_5$ function we get nearly the same turn-over magnitudes and, 
as expected \citep{sec92}, somewhat narrower dispersions.

  Within the error bars, the turn-over magnitudes and dispersions of the 
red and blue GCs are thus compatible with a standard GCLF, although the red 
GCs formally appear to have a somewhat narrower dispersion and a slightly
fainter turn-over magnitude. The fact that the red GCs tend to have
fainter turn-over magnitudes than blue ones has also been noted in other
galaxies like e.g. NGC~3115 \citep{kundu98}.  We also used
a Kolmogorov-Smirnov test \citep{lin62} to test how well the data are
actually fitted by the Gaussian and $t_5$ functions, as indicated by the
$P$ values in Table~\ref{tab:lf}.  $P$ gives the probability that the data 
are drawn from parent distributions with the specified parameters.
The blue compact clusters are very well fitted by both the Gaussian ($P=0.991$)
and the $t_5$ function ($P=0.999$), with the $t_5$ function being slightly
preferred 
over the Gaussian as was also found by \citet{sec92} for Milky Way and M31
globular clusters. For the red clusters, on the other hand, both the
Gaussian and the $t_5$ function provide a quite poor fit to the data
with $P$ values of $0.352$ and $0.529$ although the $t_5$ function is
again preferred.

  We also compared the luminosity functions of the red and blue compact 
populations directly using a two-population Kolmogorov-Smirnov test.
Such a test is independent of any assumptions about the LF.
Again leaving out clusters with $V>24$ in order to obtain as pure a sample 
of ``true'' globular clusters as possible, the $D_n$ statistics of the test 
is 0.154.  This corresponds to a 51\% probability that the two cluster 
populations have the same luminosity function. The test is thus inconclusive 
as to whether or not the two compact cluster populations have the same 
luminosity function. This remains true even if the K-S test is performed 
only on a brighter subsample of the clusters.

  The {\it extended} red objects do have a quite different luminosity 
function from a standard Gaussian (lower left panel), increasing at least 
down to $V\sim24$. 
Because of extendedness of these objects, incompleteness effects have 
already set in at $V\sim24.5$ or so (cf.\ Figs.~\ref{fig:cplot} and 
\ref{fig:cmpl2}), so it remains uncertain how their luminosity function 
behaves at fainter magnitudes. With typical effective radii of the order
of 10--15 pc (Fig.~\ref{fig:vi_fwhm}), the $V$ magnitudes of the 
extended objects are probably underestimated by $\sim 0.5$ mag
(Table~\ref{tab:apc}) although the colors remain unaffected by object
size. In any case, the differences in the luminosity and color distributions 
of extended and compact objects clearly cannot be accounted for by 
instrumental effects. 

  The lower right panel of Fig.~\ref{fig:lf} displays the combined
luminosity function of all cluster candidates. Even without performing any
statistical tests it is clear that a Gaussian or a $t_5$ function
provide a poor fit to the combined LF, mostly due to the presence 
of the extended red objects. Indeed, the maximum-likelihood fit returns a
turn-over magnitude of $\mto = -6.98^{-7.13}_{-6.85}$ for the Gaussian and
$\mto = -7.07^{-7.21}_{-6.95}$ for the $t_5$ function, half a magnitude 
fainter than for the compact clusters alone. 

  Bearing in mind that the magnitudes for extended objects measured in the
$r=3$ pixels aperture are probably too faint by $\sim 0.5$ mag 
(Sect.~\ref{sec:apc}), the true luminosity function of all objects may
look somewhat different from the lower right panel of Fig.~\ref{fig:lf}.
We added a rough correction of $-0.5$ mag to the $V$ band 
magnitudes of the extended objects in order to check how this
would affect the combined luminosity function. However, applying the same 
magnitude limit of $V=24$ to the corrected sample would then lead to a 
larger number of faint objects with uncertain sizes and photometry, 
possibly resulting in unreliable LF fits. We therefore restricted the 
LF fits of the corrected sample to objects brighter than $V=23.5$. For
a Gaussian fit we then obtained a turn-over at 
$\mto = -7.19^{-7.35}_{-6.77}$, which is somewhat closer to the value 
expected for a ``standard'' Gaussian GCLF. If we apply the same $V=23.5$ 
magnitude cut to the original {\it non-corrected} photometry, we obtain 
$\mto = -7.25^{-7.42}_{-6.85}$ for a Gaussian fit. As demonstrated by 
these experiments, the LF fits to the full cluster sample including 
extended objects are quite unstable and depend strongly on the magnitude 
cuts and various corrections. This is not too surprising, considering that 
we are attempting to fit a Gaussian function to a non-gaussian distribution.

  Until it is better understood how common such extended faint clusters 
are and what fraction of the total GC population they constitute in 
different galaxies, they could pose a potential problem for the use of
GCLFs as ``standard candles''.  However, Gaussian or $t_5$ fits 
to GCLFs should still be relatively secure if constrained to objects 
brighter than about 1 mag below the turn-over.

\section{Discussion}

  In summary, NGC~1023 appears to contain at least three identifiable 
cluster populations: 1) blue globular clusters with a standard Gaussian 
luminosity function and with an average $\reff \sim 2$ pc. The mean 
color is $\langle\vi\rangle_0 = 0.92$. 2) red globular clusters 
($\langle\vi\rangle_0 = 1.15$) with smaller sizes ($\reff \sim 1.7$ pc), 
whose LF may have a somewhat fainter turn-over
magnitude and a slightly narrower dispersion. 3) A population of faint 
extended sources with $\reff \sim 10 - 15$ pc ($\reff \sim 7 - 10$ pc 
when using MOFFAT15 model profiles), which are predominantly red.

\subsection{The classical globular clusters}

  Concerning sizes and color distributions of the normal compact clusters, 
the GCS of NGC~1023 appears to be quite similar to what has been observed 
in a number of other early-type galaxies such as NGC~4472, M87 and NGC~3115
\citep{kundu98,kun99,puz99}.  Deviation from unimodality in the NGC~1023 
cluster colors was already hinted at by \citet{geb99} and is now very 
clearly confirmed by our deeper data.  The fact that red GCs tend to have 
smaller effective radii than their blue counterparts has also been observed 
in the above mentioned galaxies, but the small distance of NGC~1023 and
our deep HST data add further confidence to this result.

  The overall characteristics of the cluster systems in a number of 
early-type galaxies thus appear to be quite similar, providing support to the 
idea that the same mechanisms governed formation of the globular cluster 
systems in a wide variety of galaxies. In particular, any scenario 
that attempts to explain the presence of bimodal color distributions must
account for the presence of this phenomenon in cDs as well as lenticulars 
and other galaxy types. 

  Ultimately, one also needs to understand the size difference
between metal-poor and metal-rich clusters -- perhaps this is evidence
for different physical conditions in the parent clouds, or maybe it
indicates that the different cluster populations have been subject to 
different destruction processes, e.g.\ because of different kinematics.
Obviously, a better understanding of the physics of cluster formation 
and evolution is 
needed in order to address the first possibility, while insight into the 
kinematics of cluster sub-populations may be gained by spectroscopy of large
cluster samples. It is worth pointing out, though, that all theoretical
studies so far indicate that the half-light radius for any individual
cluster is a very stable quantity and does not change much over the
lifetime of the cluster \citep{spi87,mey97}.  If the size distributions
of red and blue clusters in NGC~1023 (and other galaxies) originated from 
the same parent distribution, quite substantial differences in the
destruction mechanisms and/or dynamical evolution of the
two subpopulations would have been necessary to make their present-day size 
distributions differ as much as is observed. At the same time, these 
mechanisms would have to act in such a way that the blue and red GCs would 
still retain roughly similar luminosity functions, although \citet{ves00} 
has recently argued that the GCLF is relatively stable against dynamical
evolution.

Alternatively, different
cluster sizes would be the result of different physical conditions at
the time of cluster formation. Our understanding of cluster formation is
currently in a rather rudimentary state, but in the case of the Milky
Way, \citet{mac99} has suggested that the observed correlation between 
sizes of globular clusters and galactocentric distance could be explained
by the progenitor clouds closer to the center having higher binding
energies.  How this relates to the size differences between red and blue 
GC populations in NGC~1023 and other early-type galaxies is not entirely 
clear though, especially since no correlation between galactocentric 
distance and cluster sizes appears to exist in galaxies like NGC~4472
\citep{puz99}. One could speculate that, since the red (metal-rich) population
is likely to have formed somewhat later than the blue one, the key issue 
may here be a {\it temporal} rather than {\it spatial} change in the cluster 
forming environment. 

\subsection{A third cluster population?}

  A major peculiarity about the NGC~1023 cluster system is the presence
of a number of fainter objects with much larger average effective 
radii than the ``genuine'' globular clusters. Most of these objects are 
quite red with colors comparable to the red GC population, $\viz \sim 1.2$.
Because the comparison field does not contain similar objects, and because
of the alignment of their spatial distribution with the isophotes of
NGC~1023 itself, we feel confident that these objects are indeed star
clusters in NGC~1023. A chance alignment with e.g.\ a distant galaxy 
cluster appears to be highly unlikely, not only because of the low
probability of such an alignment by itself, but even more so because the
background cluster would have to have an annular shape.
  
  The red colors of these objects are probably real and not due to e.g.\
reddening by interstellar absorption. First, reddening is not expected to 
be much of an issue in a galaxy of
this type. This is confirmed by the absence of any significant FIR
emission -- the 12$\mu$ flux as measured by the IRAS satellite is only 
0.16 Jy \citep{ric88} and in the remaining IRAS bands only upper 
limits exist, so the presence of any significant amount of dust is ruled out.
Furthermore, although the NGC~1023 system does possess significant amounts
of HI gas, most of it is located far outside the main body of NGC~1023.
Finally, interstellar absorption could not account for the {\it extendedness} 
of the faint clusters.

  What is the nature of the faint extended clusters? Generally, they
are too bright to be normal open clusters. For ages $> 1$ Gyr, the 
\citet{lyn87} catalog lists no open clusters in the Milky Way brighter 
than $M_V \sim -4$.  Furthermore, open clusters are not particularly 
large, with mean and median radii of 2.7 pc and 1.9 pc, respectively 
\citep{lyn82}.  \citet{hod79} finds larger sizes for open clusters in 
M31, but notes that his sample is biased towards large clusters because 
of resolution limits. Hence, the extended clusters in NGC~1023 are too 
bright and too extended to be normal open clusters.

  The $V$ range spanned by the extended objects ($23 - 24$) corresponds 
to absolute magnitudes 
between $-6$ and $-7$, well within the range spanned by {\it globular} 
clusters in our own and other galaxies and nearly reaching the peak of 
the standard GCLF. Our magnitudes for the extended objects are actually 
likely to be too {\it faint} by up to several tenths of a magnitude 
because of a larger fraction of light falling outside the $r=3$ 
aperture (Sect.~\ref{sec:apc}).  The closest counterparts in the Milky Way 
would be the Palomar-type globular clusters found in the outer part of 
the Galactic halo. Actually, most globular clusters in the Milky Way with 
{\it core} radii $R_c > 3$ pc are known to be systematically fainter than 
globular clusters with $R_c < 3$ pc, very few of them being brighter than 
$M_V = -7$ \citep{van83,van96}. However, the extended clusters in
NGC~1023 appear to be relatively {\it metal-rich} whereas the extended
clusters in the outer halo of the Milky Way are metal-poor.

  Concerning the origin of the extended red clusters, we may speculate that
they could be related to accretion of satellite galaxies. In particular, 
ground-based images reveal the presence of two bright blue condensations 
near the center of the companion NGC~1023A which, unfortunately, were not 
included on our WFPC2 pointings. These blue objects were already noted
by \citet{dav84} who suggested they might be supergiant stars, but they 
are clearly extended and ground-based photometry and spectroscopy suggest
ages on the order of $\sim 0.5$ Gyr (Larsen et al., in preparation). 
Therefore NGC~1023A must have had an active star formation history 
until quite recently and a number of clusters could conceivably have been
stripped from the companion and ended up in the disk of NGC~1023 during
the last few 100 Myr or so. If NGC~1023A used to be a cluster--rich LMC--like 
galaxy, this scenario may not be entirely unlikely. One problem with this
scenario is the red colors of the extended clusters, indicating 
relatively high metallicities. This could, perhaps, be explained if
the extended clusters were {\it formed} in the disk of NGC~1023 when sufficient
gas was still available, possibly triggered by interaction processes, rather 
than simply being accreted.

  It would be highly desirable to shed more light on these questions 
with observations of similar objects in other galaxies.
However, to our knowledge no extended, faint red clusters like those in 
NGC~1023 have previously been identified in any other early-type galaxy. 
The main point, of course, is that these objects are relatively {\it faint} 
and thus difficult to detect, and that high angular resolution is needed 
to obtain any information on cluster sizes. Therefore, only deep 
HST observations of nearby ($\la 10$ Mpc) galaxies would reveal such 
objects and they would, so far, have escaped detection even in HST 
studies of Virgo cluster galaxies. Until now, only one other early-type 
galaxy close enough for detection of extended clusters similar to those 
in NGC~1023 has had sufficiently deep HST photometry, namely 
NGC~3115 \citep{kundu98}.  However, \citet{kundu98} did not report any
such objects in their study of this galaxy. In order to check if they
might have been overlooked, we re-examined the archive data and found
that extended faint objects are indeed absent in NGC~3115.

\section{Summary and conclusions}

   We have obtained deep WFPC2 photometry for star clusters in the
nearby lenticular galaxy NGC~1023, reaching two magnitudes below the
expected turn-over of the globular cluster luminosity function.
Our \vi\ photometry reveals a very clearly bimodal color distribution
with peaks at $\viz = 0.92$ and $\viz = 1.15$, respectively. We also
identified a number of red objects with larger effective radii than 
the brighter GCs and a luminosity function that rises at least down
to $M_V\sim-6$ (assuming the objects are at the distance of NGC~1023). 
A comparison field located about 2 degrees from NGC~1023
shows no such objects. Moreover, the spatial distribution of faint
extended objects is well aligned with the isophotes of NGC~1023, 
indicating a very high probability that they are star clusters in the 
disk of NGC~1023.

  The blue GC population has a luminosity function that is consistent 
at the $99\%$ confidence level with a ``standard'' Gaussian GCLF with 
$\mto = -7.58^{-7.72}_{-7.36}$ and $\sigma_V = 1.12^{1.33}_{1.03}$.
A $t_5$ function provides an even better fit, with 
$\mto = -7.56^{-7.70}_{-7.37}$ and $\sigma_V = 0.99^{1.20}_{0.90}$ at
the $99.9$\% confidence level.  The average effective radius of the blue
GCs is $\sim 2$ pc, comparable to (but perhaps slightly smaller than) that
seen in the Milky Way and other galaxies.  The LF of the red GC 
population may have a somewhat narrower dispersion and a peak magnitude
about 0.2 mag fainter than for the blue GC population, but is not 
very well fitted by either a Gaussian or a $t_5$ function. On 
the average, the brighter red GCs have $\reff\sim 1.7$ pc, or about
15\% smaller than for the blue GCs.

  The spatial, color and luminosity distributions of {\it extended} clusters 
with $\reff > 7$ pc are clearly different from those of compact ``genuine'' 
globulars.  We suggest that the faint extended red clusters may have 
their origin in accretion processes with companion galaxies, similar to
the ongoing interaction with the dwarf NGC~1023A.  Because of the extended 
red clusters, the combined luminosity function of all cluster candidates 
in NGC~1023 deviates significantly from a standard Gaussian GCLF. If 
such objects are common in other galaxies, this could potentially lead 
to problems in using the GCLF as a standard candle.

\acknowledgments

This work was supported by HST grant GO.06554.01-95A,
National Science Foundation grant number
AST9900732 and Faculty Research funds from the University of California,
Santa Cruz. We thank Duncan Forbes and John Huchra for useful comments 
and suggestions, Carl Grillmair and Ken Freeman for their support, 
and the anonymous referee for a detailed and constructive report which 
helped significantly improve the paper.

\onecolumn

\begin{minipage}{155mm}
\epsfxsize=12cm
\epsfbox{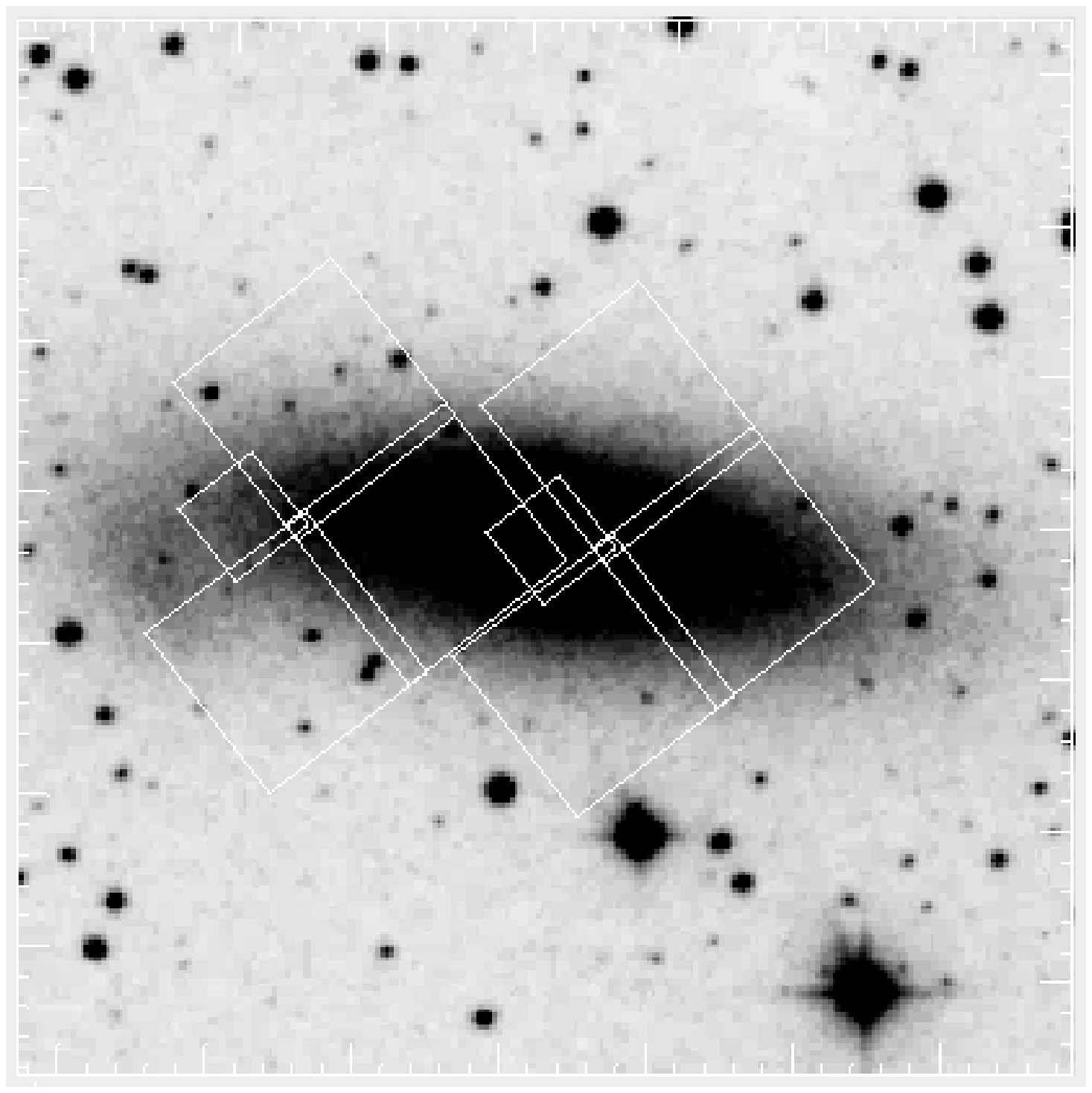}
\figcaption[larsen.fig1.ps]{\label{fig:pointings} \sl
  Our two WFPC2 pointings on NGC~1023, with north up and east to the left. 
The dwarf companion NGC~1023A is seen as an extension to the E.}
\end{minipage}

\begin{minipage}{155mm}
\epsfxsize=12cm
\epsfbox{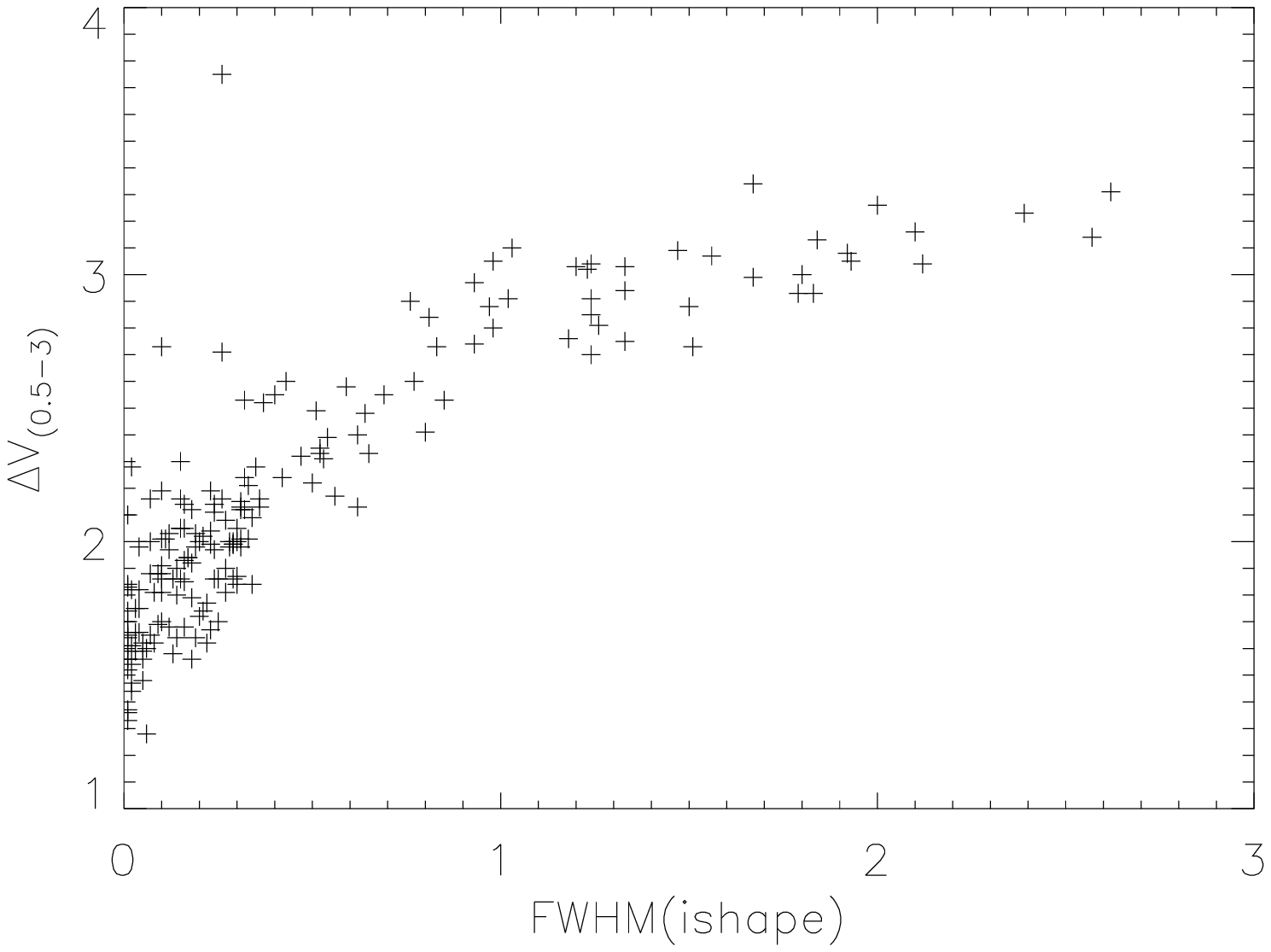}
\figcaption[larsen.fig2.ps]{\label{fig:delta} \sl
  The $\Delta_{0.5-3}$ index, defined as the difference between magnitudes
measured through $r=3$ and $r=0.5$ pixel apertures on an F555W image,
plotted against FWHM estimates by \ishape\ for objects brighter than 
$V=24$.}
\end{minipage}

\begin{minipage}{155mm}
\epsfxsize=7cm
\epsfbox{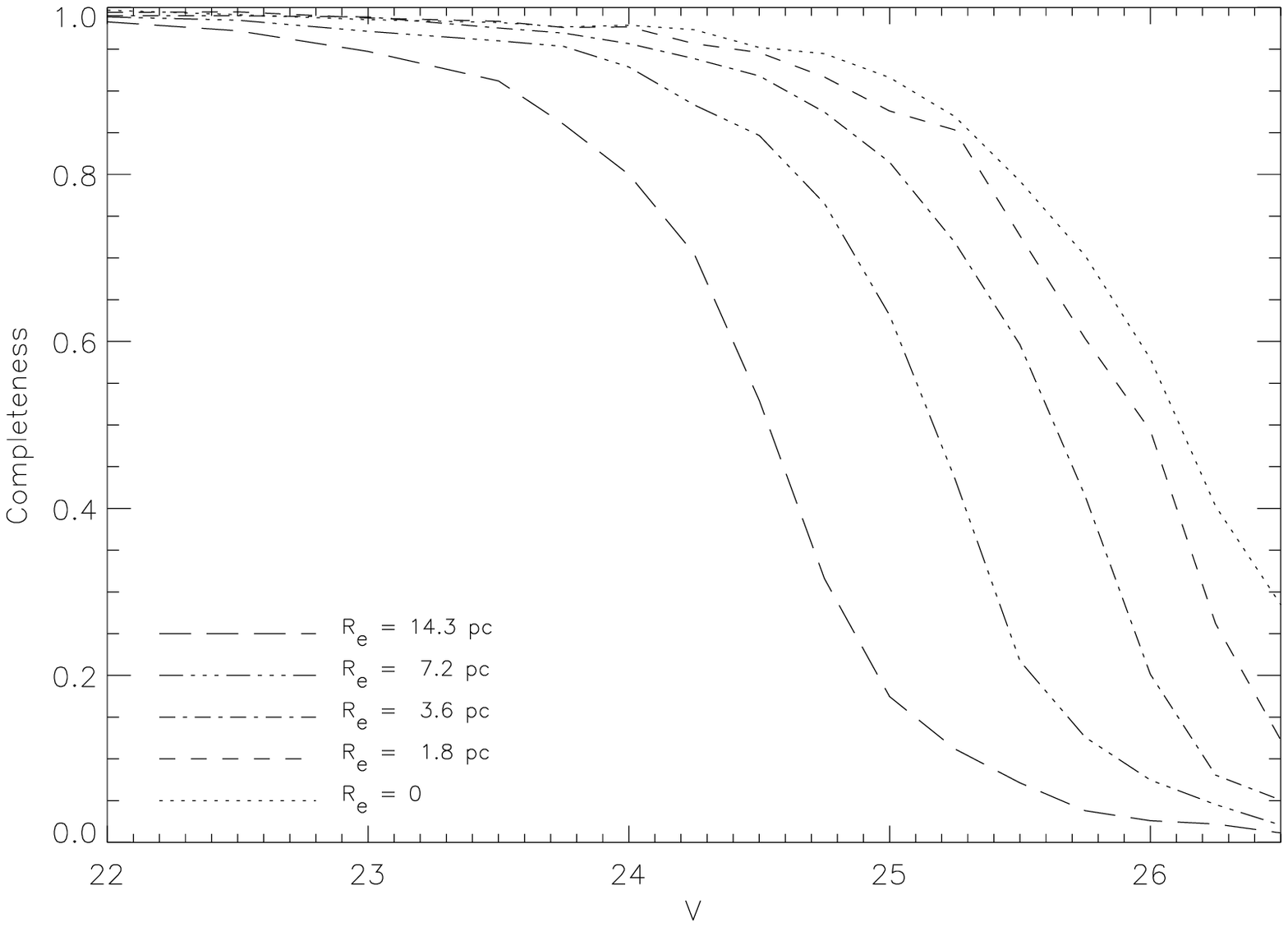}
\epsfxsize=7cm
\epsfbox{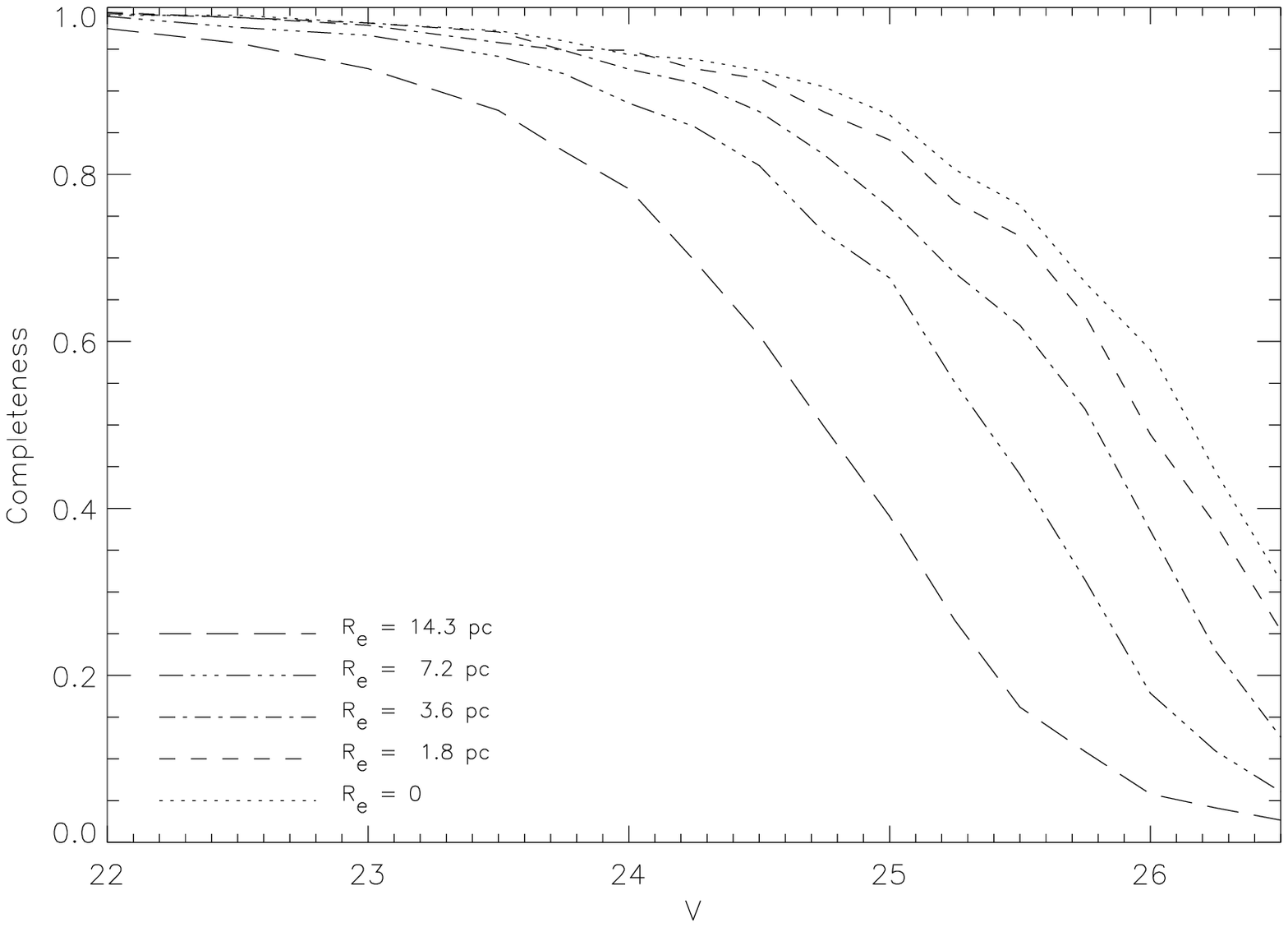}
\figcaption[larsen.fig3a.ps,larsen.fig3b.ps]{\label{fig:cplot} \sl
  Completeness functions for the WF chips in the
  center (left) and east (right) pointings, shown for cluster
  FWHM values of 0, 0.25, 0.50, 1.0 and 2.0 pixels (corresponding to
  \reff\ = 0, 1.8, 3.6, 7.2 and 14.2 pc, respectively). As this figure
  demonstrates, the completeness corrections depend strongly on 
  cluster size.}
\end{minipage}

\begin{minipage}{155mm}
\epsfxsize=12cm
\epsfbox{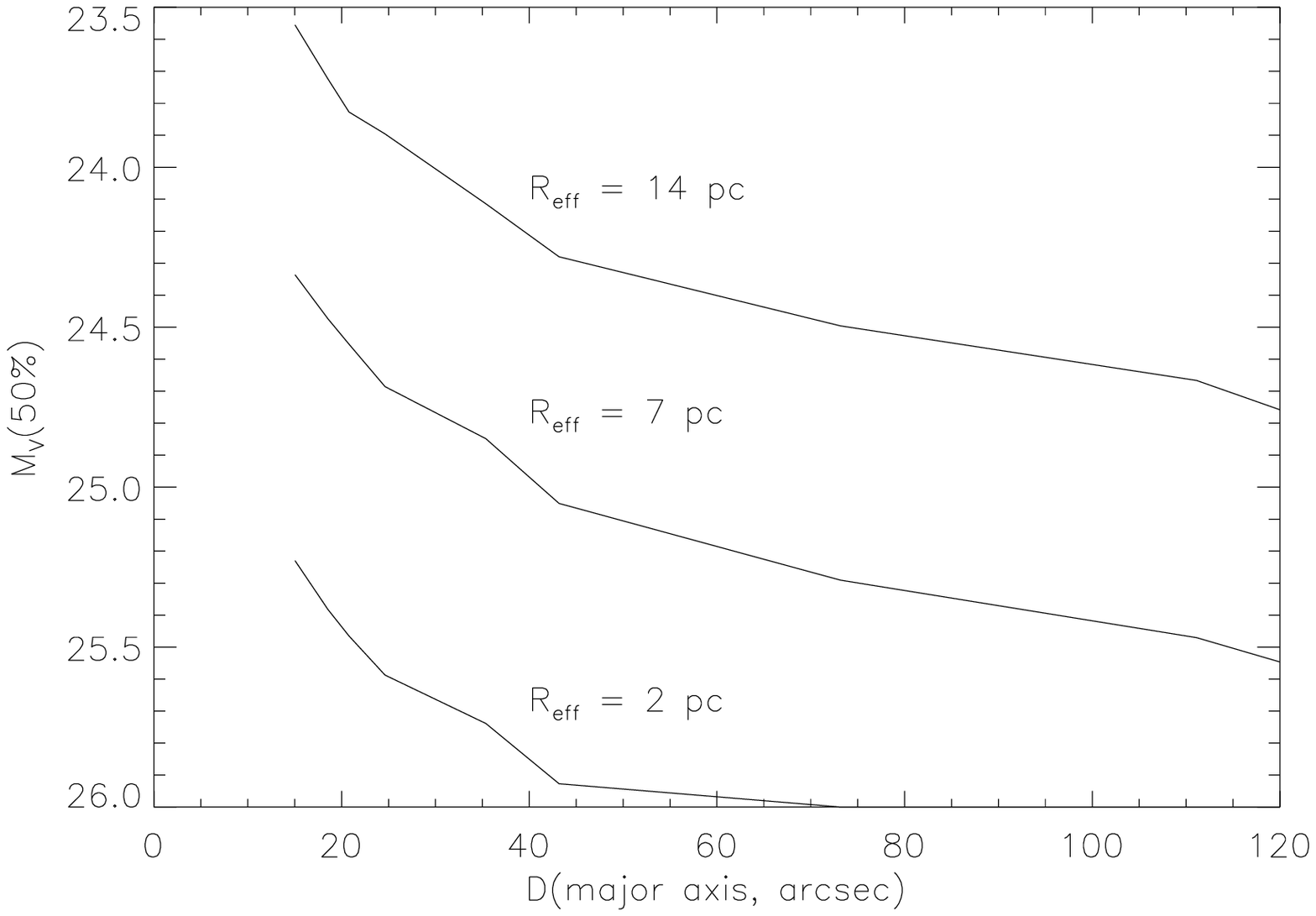}
\figcaption[larsen.fig4.ps]{\label{fig:cmpl2}\sl 50\% completeness limits
as a function of distance from the center of NGC~1023 along the major 
axis. Curves are shown for three different object sizes: $\reff =$ 2 pc, 
7 pc and 14 pc.}
\end{minipage}

\begin{minipage}{155mm}
\epsfxsize=12cm
\epsfbox{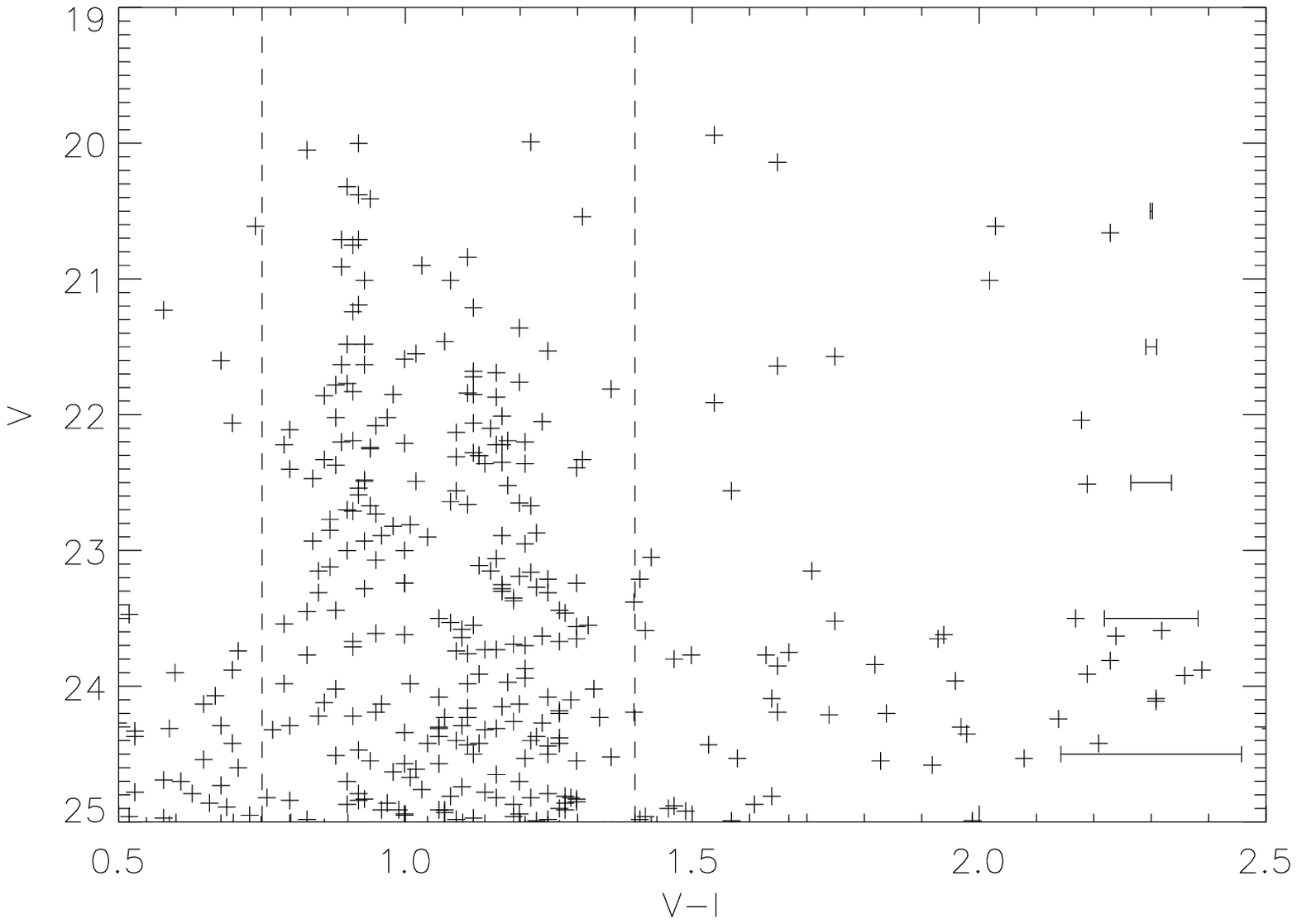}
\figcaption[larsen.fig5.ps]{\label{fig:vi_v_all} \sl
  A $\vi , V$ color-magnitude diagram for all objects detected in 
  the two HST pointings on NGC~1023. Two peaks in the color distribution,
  at $\vi = 0.92$ and at $\vi = 1.19$ are readily distinguished. Photometric
  errors are indicated by the error bars at $\vi = 2.3$.
  }
\end{minipage}

\begin{minipage}{155mm}
\epsfxsize=12cm
\epsfbox{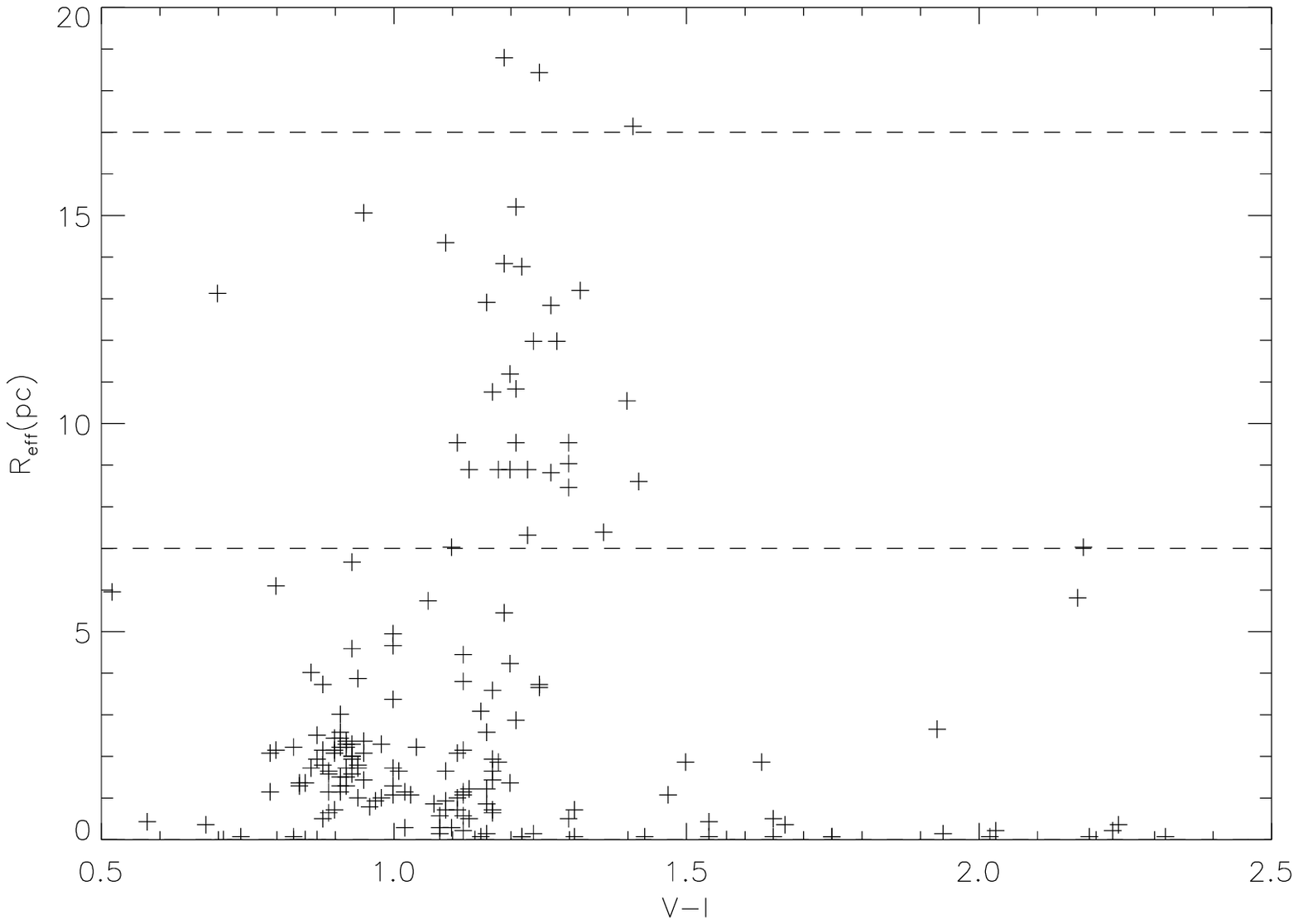}
\figcaption[larsen.fig6.ps]{\label{fig:vi_fwhm} \sl
  Effective radii (in pc) as a function of \vi\ color for objects
  brighter than $V=24$.  Cluster sizes were measured using the \ishape\ 
  algorithm (Larsen, 1999) assuming King $c=30$ profiles.  Dashed lines 
  mark the cuts applied in the selection of regular globular cluster 
  candidates and ``fuzzy'' disk clusters.}
\end{minipage}

\begin{minipage}{155mm}
\epsfxsize=12cm
\epsfbox{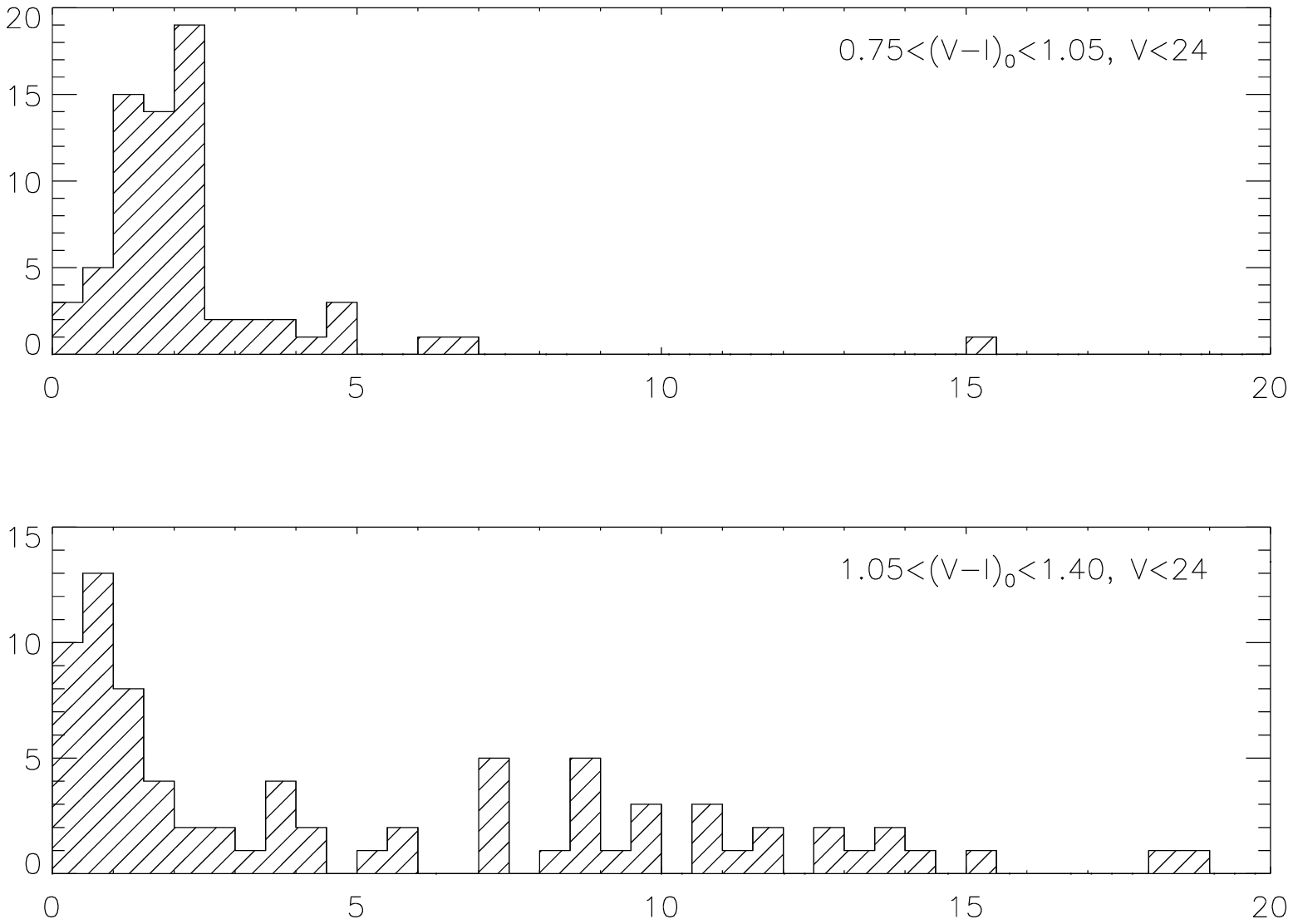}
\figcaption[larsen.fig7.ps]{\label{fig:szdist} \sl
  The size distribution for blue (top) and red (bottom) objects with
$V<24$. The majority of red objects are somewhat smaller than the blue
ones, but the size distribution of red objects has a tail extending up 
to $\reff \sim 15$ pc.
}
\end{minipage}

\begin{minipage}{155mm}
\epsfxsize=12cm
\epsfbox{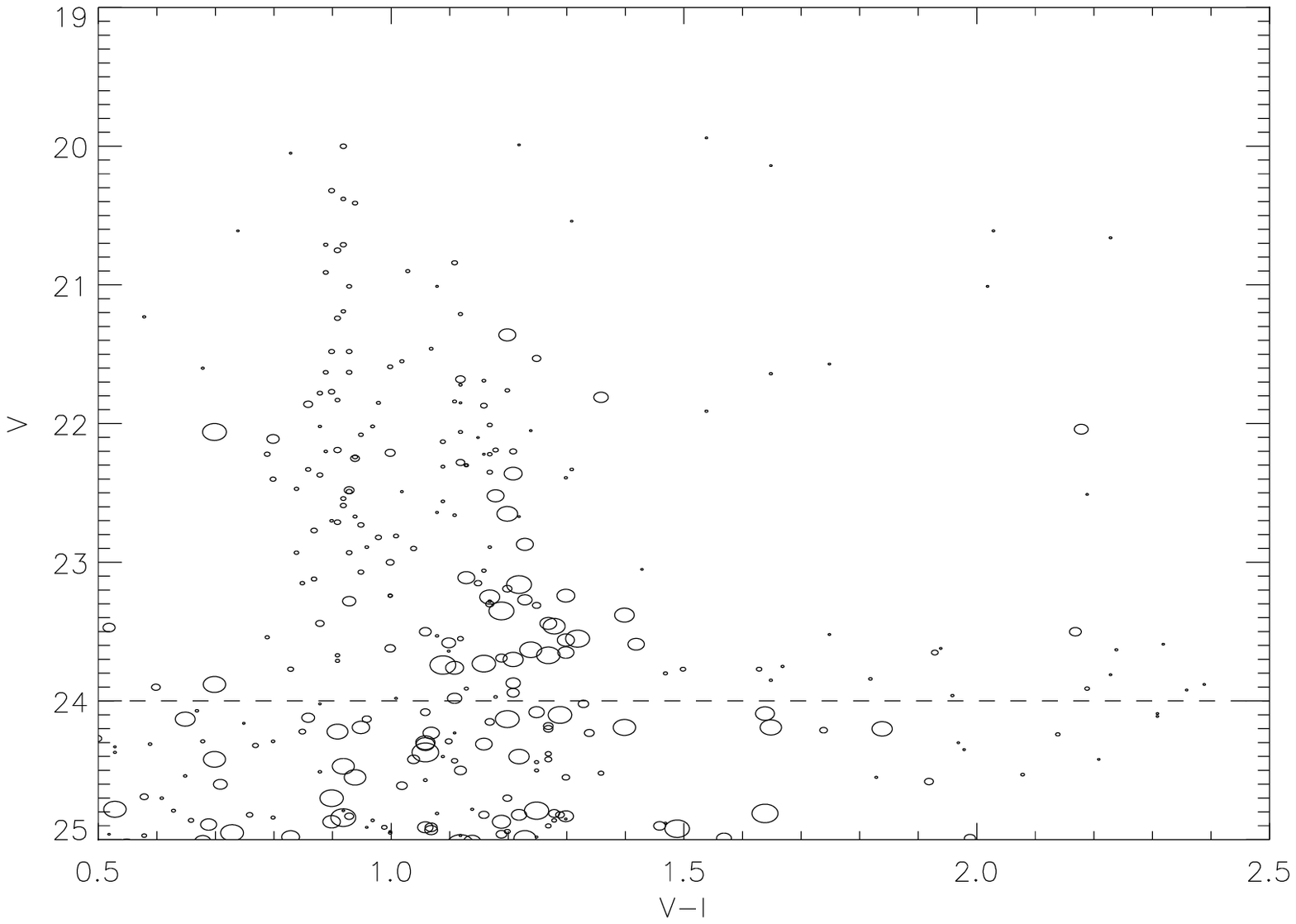}
\figcaption[larsen.fig8.ps]{\label{fig:pops} \sl
  Color-magnitude diagram with symbol sizes representing the
  object sizes measured by \ishape . Note the concentration of faint
  extended objects near $V=23.5$ and $\vi = 1.2$. The horizontal line
  indicates the magnitude limit above which cluster sizes are accurate
  to better than $\pm$ 30\% .
  }
\end{minipage}

\begin{minipage}{155mm}
\epsfxsize=12cm
\epsfbox{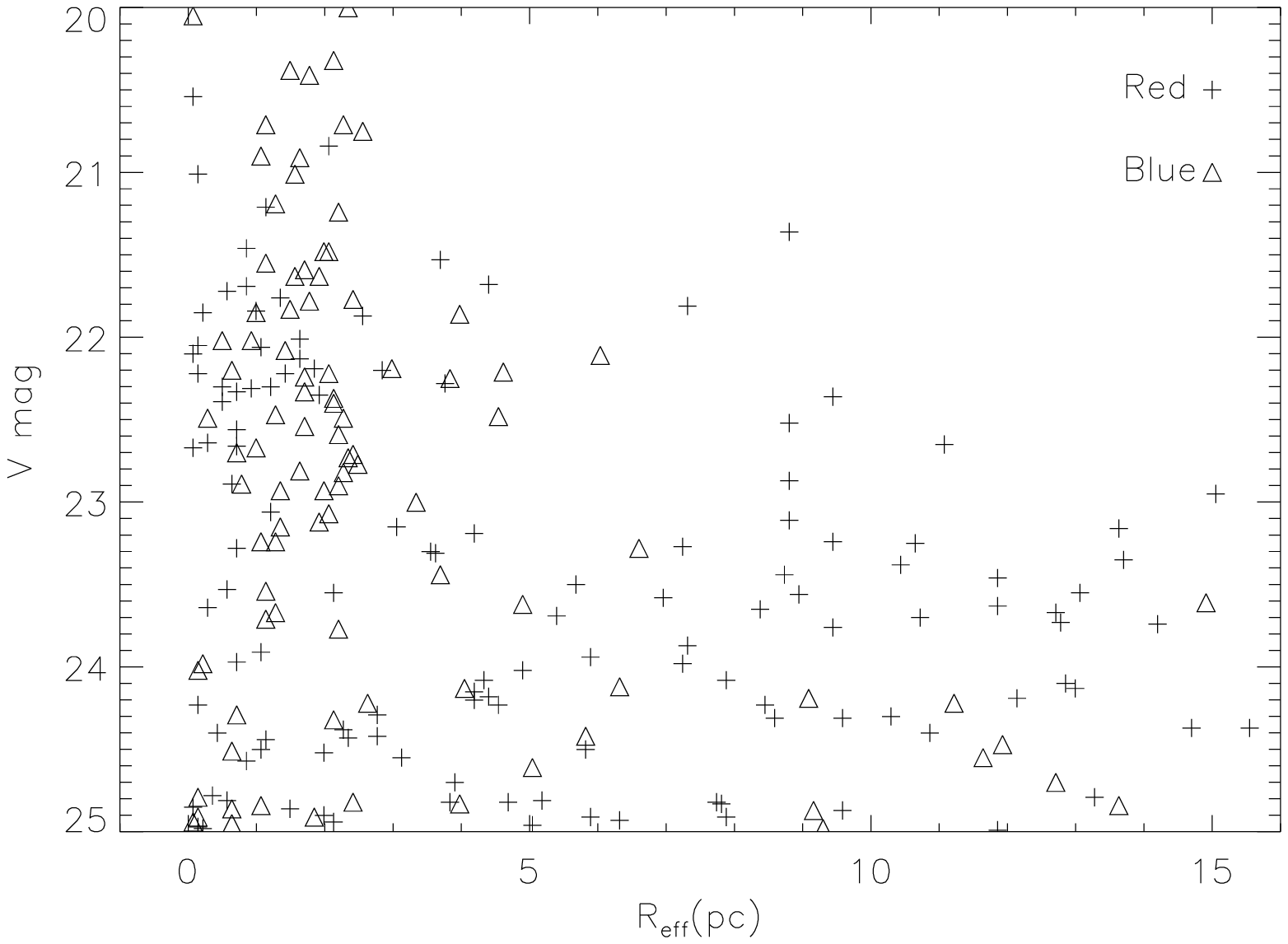}
\figcaption[larsen.fig9.ps]{\label{fig:sz_v} \sl
  $V$ magnitudes versus cluster sizes. Red clusters ($\vi > 1.05$)
are shown with plus markers and blue clusters with triangles. 
Qualitatively, this plot shows a striking resemblance to Fig. 1 of 
\citet{van96} although extended clusters appear to be much more 
numerous in NGC~1023.
}
\end{minipage}

\begin{minipage}{155mm}
\epsfxsize=12cm
\epsfbox{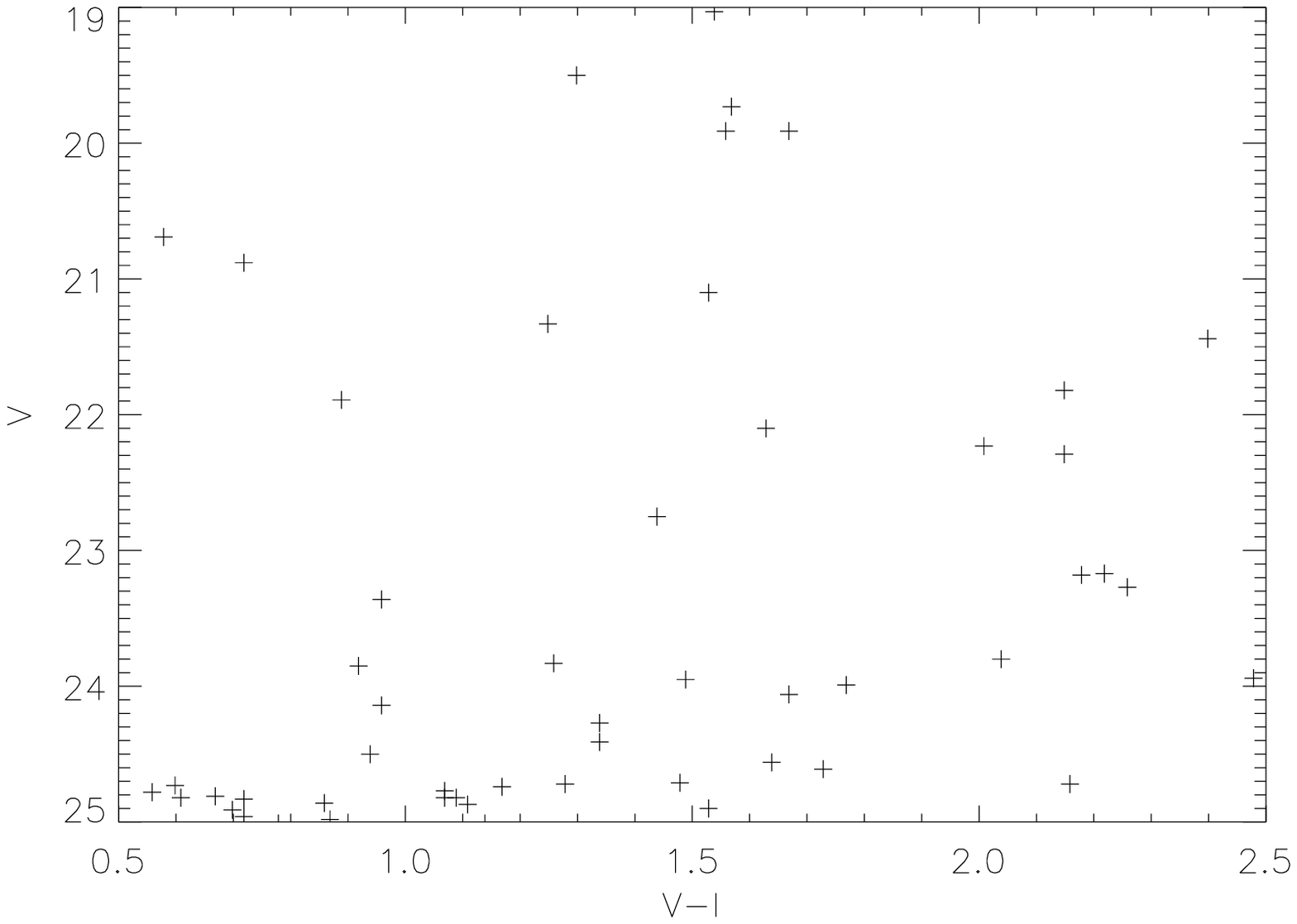}
\figcaption[larsen.fig10.ps]{\label{fig:cmp} \sl
  Color-magnitude diagram for the comparison field located about 2 degrees
from NGC~1023. 
}
\end{minipage}

\begin{minipage}{155mm}
\epsfxsize=14cm
\epsfbox{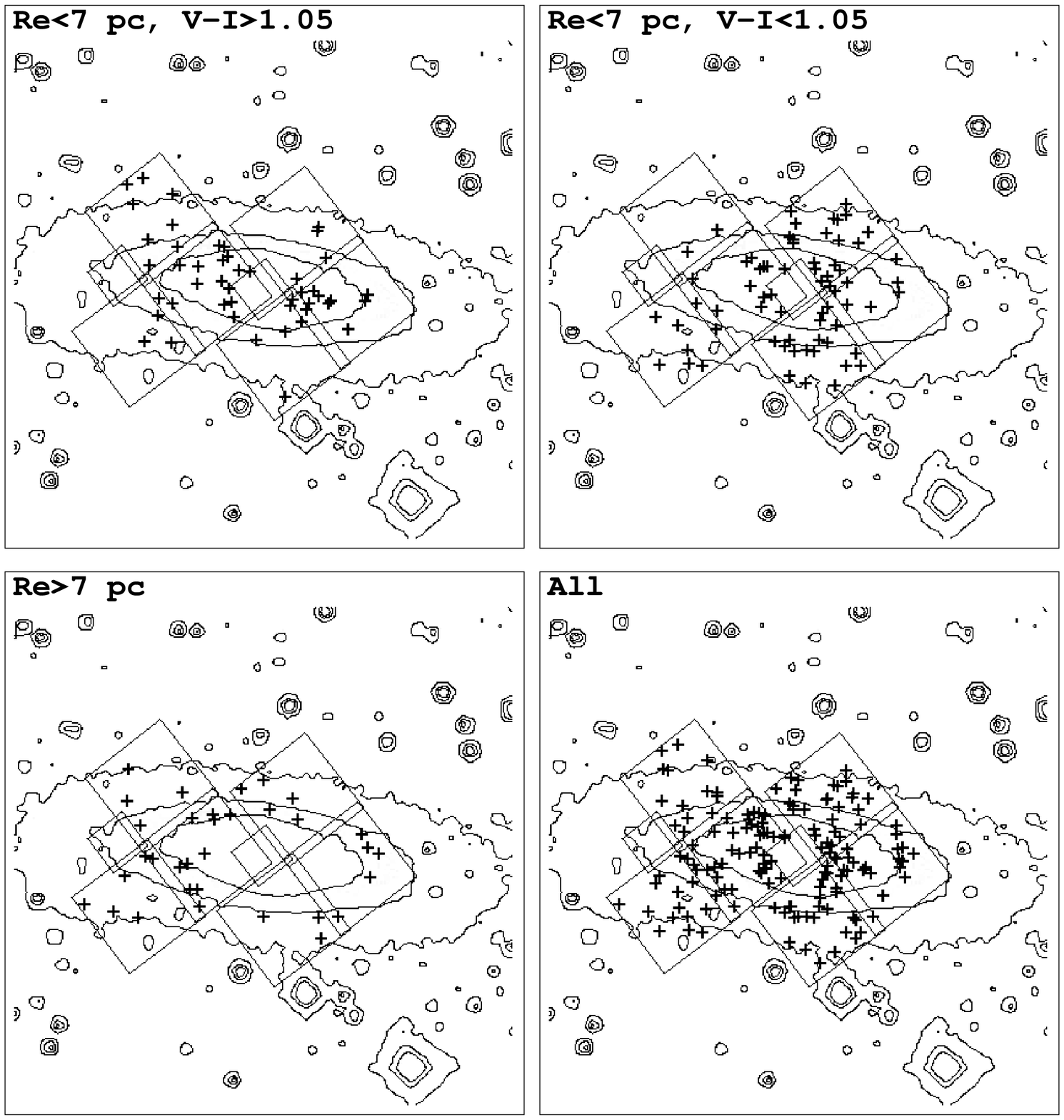}
\figcaption[larsen.fig11.ps]{\label{fig:dist} \sl
  The spatial distribution of different object types.}
\end{minipage}

\begin{minipage}{155mm}
\epsfxsize=12cm
\epsfbox{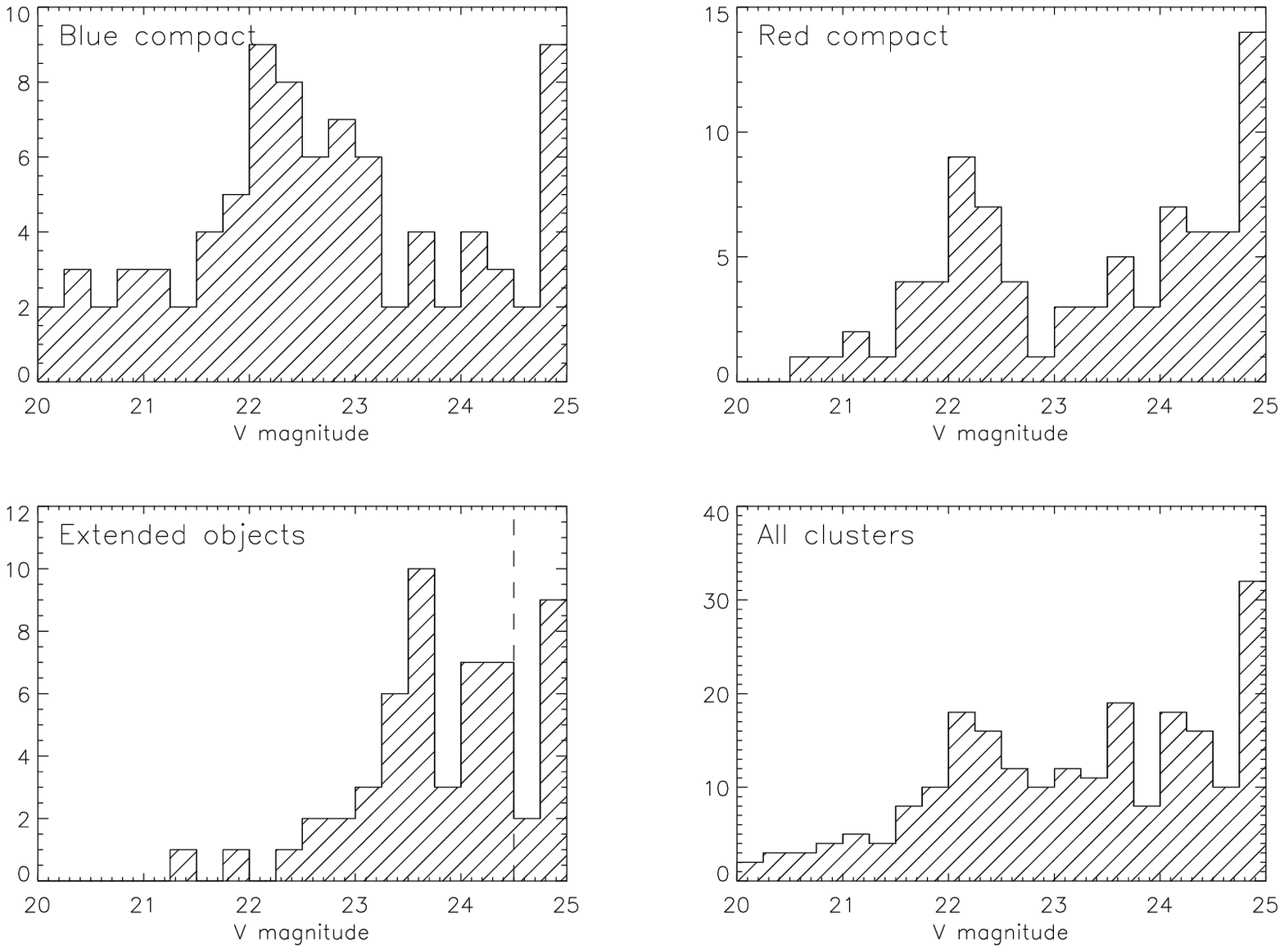}
\figcaption[larsen.fig12.ps]{\label{fig:lf} \sl
  Luminosity functions for the 3 cluster populations in
  NGC~1023. Approximate 50\% completeness limits are indicated by the 
  dashed lines. The brighter 50\% limit for the extended clusters is due
  to their larger sizes. For the compact objects (upper panels) the
  50\% completeness limits are outside the plotted range.}
\end{minipage}

\begin{minipage}{155mm}
\epsfxsize=7cm
\epsfbox{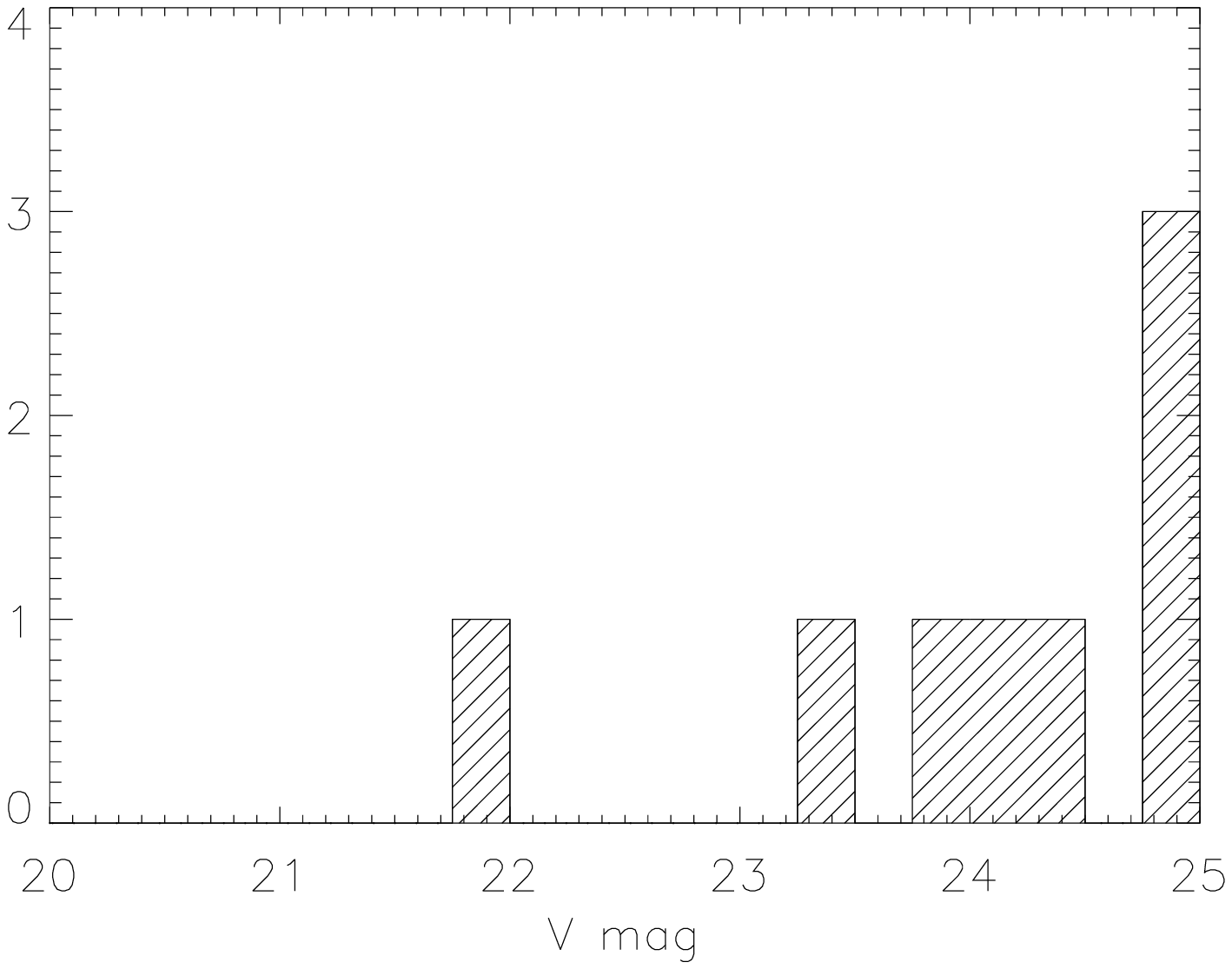}
\epsfxsize=7cm
\epsfbox{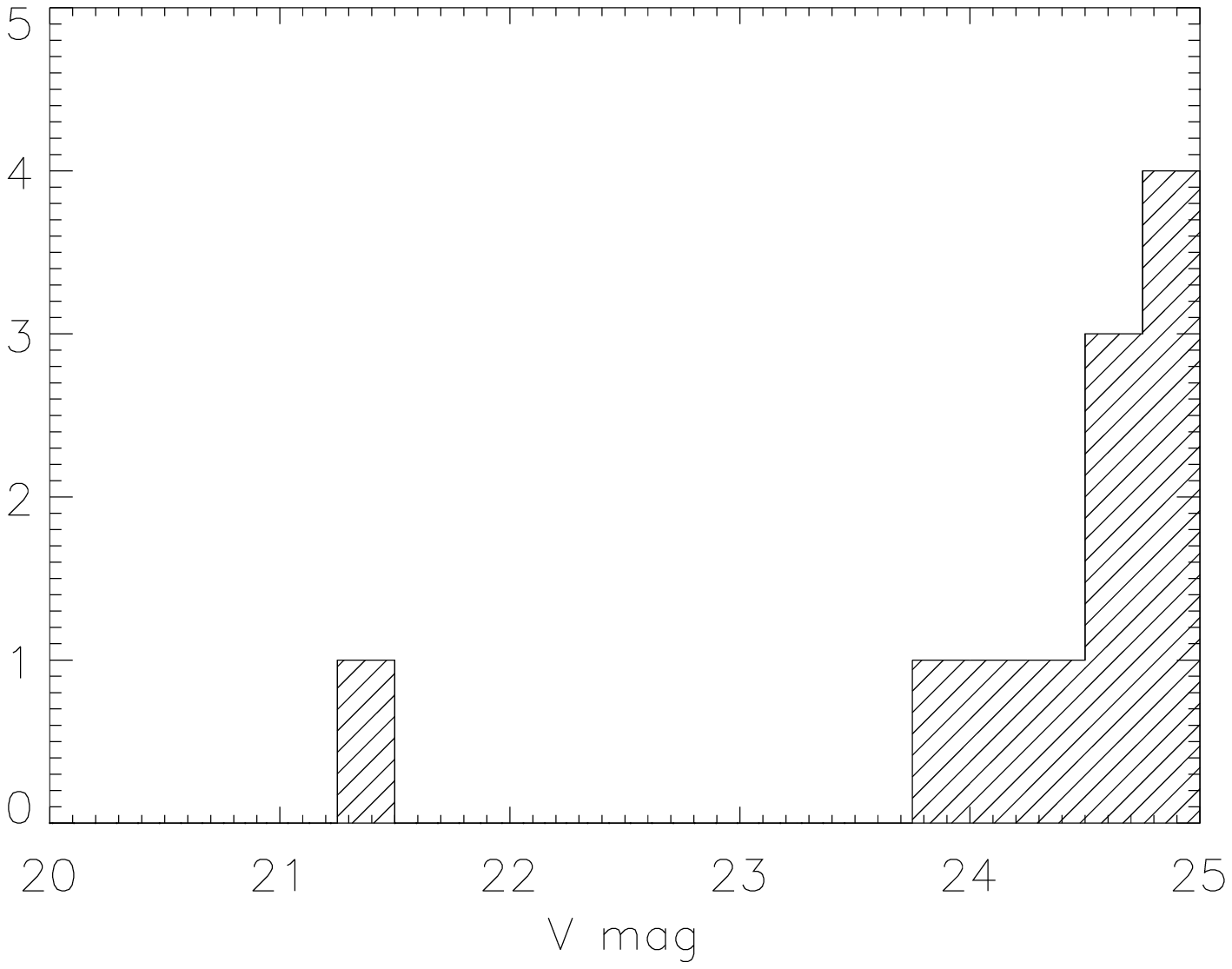}
\figcaption[larsen.fig13b.ps,larsen.fig13a.ps]{\label{fig:lf_cmp} \sl
  Luminosity functions for blue (left) and red (right) objects in the
comparison field.
}
\end{minipage}

\newpage

\begin{table}
\begin{center}
\caption{Aperture corrections for different model profiles convolved
with the \tinytim\ PSF and diffusion kernel. The FWHM
are in pixel units. Equivalent $\reff$ values are given for a distance of 
9.9 Mpc.  The notation $5\rightarrow3$ refers to the
aperture correction from an $r=5$ pixels to an $r=3$ pixels aperture.
\label{tab:apc}
}
\begin{tabular}{rrrrrrrr}
\tableline\tableline
 Profile & $\reff$(pc) &
           \multicolumn{3}{c}{F555W} & \multicolumn{3}{c}{F555W$-$F814W} \\
         & & $5\rightarrow3$ & $5\rightarrow2$ & $30\rightarrow5$
           & $5\rightarrow3$ & $5\rightarrow2$ & $30\rightarrow5$ \\
\tableline
\multicolumn{1}{l}{King $c=30$} & \multicolumn{7}{c}{} \\
FWHM= 0.00  & 0.0 & $-0.060$ & $-0.178$ & $-0.09$ & 0.015 & 0.031 & 0.01 \\
      0.10  & 0.7 & $-0.061$ & $-0.185$ & $-0.09$ & 0.016 & 0.031 & 0.01 \\
      0.25  & 1.8 & $-0.079$ & $-0.248$ & $-0.09$ & 0.015 & 0.030 & 0.01 \\
      0.50  & 3.5 & $-0.157$ & $-0.406$ & $-0.11$ & 0.014 & 0.027 & 0.01 \\
      1.00  & 7.1 & $-0.276$ & $-0.625$ & $-0.24$ & 0.013 & 0.024 & 0.01 \\
      2.00  &14.2 & $-0.426$ & $-0.899$ & $-0.52$ & 0.011 & 0.020 & 0.01 \\
\multicolumn{1}{l}{MOFFAT15} & \multicolumn{7}{c}{} \\
FWHM= 0.00  & 0.0 & $-0.060$ & $-0.178$ & $-0.09$ & 0.015 & 0.031 & 0.01 \\
      0.10  & 0.5 & $-0.071$ & $-0.205$ & $-0.10$ & 0.016 & 0.030 & 0.01 \\
      0.25  & 1.4 & $-0.093$ & $-0.253$ & $-0.12$ & 0.015 & 0.029 & 0.01 \\
      0.50  & 2.7 & $-0.127$ & $-0.329$ & $-0.16$ & 0.014 & 0.028 & 0.01 \\
      1.00  & 5.4 & $-0.196$ & $-0.481$ & $-0.23$ & 0.015 & 0.027 & 0.01 \\
      2.00  &10.8 & $-0.335$ & $-0.758$ & $-0.38$ & 0.011 & 0.022 & 0.01 \\
\tableline
\end{tabular}
\end{center}
\end{table}

\begin{table}
\begin{center}
\caption{Results of maximum-likelihood fits of Gaussian and $t_5$ 
functions to the luminosity functions of red and blue globular clusters.
The errors in \mto\ do not include the $\pm0.14$ mag uncertainty on the
distance modulus.
$P$ is the Kolmogorov-Smirnov probability that the data are drawn from a 
parent distribution with the specified parameters.
\label{tab:lf}
}
\begin{tabular}{lcccccc}
\tableline\tableline
         &  \multicolumn{3}{c}{Gaussian} & \multicolumn{3}{c}{$t_5$} \\
         & \mto\ & $\sigma_V$ & $P$      & \mto\ & $\sigma_V$ & $P$ \\
\tableline
Blue  ($0.75<\vi <1.05$) & 
         $-7.58^{-7.72}_{-7.36}$ & $1.12_{1.03}^{1.33}$ & 0.991 &
         $-7.56^{-7.70}_{-7.37}$ & $0.99_{0.90}^{1.20}$ & 0.999 \\
Red ($1.05<\vi <1.40$) & 
         $-7.37^{-7.50}_{-7.09}$ & $0.97_{0.89}^{1.25}$ & 0.352 &
         $-7.42^{-7.56}_{-7.19}$ & $0.86_{0.78}^{1.12}$ & 0.529 \\
Red + Blue             & 
         $-7.48^{-7.59}_{-7.32}$ & $1.07_{1.00}^{1.24}$ & 0.906 &
         $-7.49^{-7.60}_{-7.37}$ & $0.94_{0.86}^{1.08}$ & 0.943 \\
Red + Blue + Extd      & 
         $-6.98^{-7.13}_{-6.85}$ & $1.27_{1.18}^{1.36}$ & 0.695 & 
         $-7.07^{-7.21}_{-6.95}$ & $1.09_{1.00}^{1.19}$ & 0.637 \\
\tableline
\end{tabular}
\end{center}
\end{table}

\end{document}